\documentclass[pdftex,twocolumn,epjc3]{svjour3}

\pdfoutput=1
\usepackage{graphics, color}
\usepackage{lineno}
\usepackage{multirow}
\usepackage{subfigure}
\usepackage{array}
\usepackage{xspace}
\usepackage{amsmath}

\RequirePackage[T1]{fontenc}
\smartqed  % flush right qed marks, e.g. at end of proof
\RequirePackage{graphicx}
\RequirePackage{mathptmx} % use Times fonts if available on your TeX system
\RequirePackage{flushend}
\RequirePackage[numbers,sort&compress]{natbib}
\RequirePackage[colorlinks,citecolor=blue,urlcolor=blue,linkcolor=blue]{hyperref}

\newcommand{\dama}{DA\-MA/LI\-BRA\xspace}
\newcommand{\nai}{\mbox{NaI}\xspace}
\newcommand{\naitl}{\mbox{NaI(Tl)}\xspace}

\newcommand{\kev}{\mbox{keV\kern-1.5pt \textsubscript{{ee}}}\xspace}
\newcommand{\kevr}{\mbox{keV\kern-1.5pt \textsubscript{{nr}}}\xspace}
\newcommand{\mev}{\mbox{MeV\kern-1.5pt \textsubscript{{ee}}}\xspace}
\newcommand{\dru}{\mbox{cpd/kg/\kev}\xspace}

\newcommand{\pop}{Proof-of-Principle\xspace}
\newcommand{\dmm}{dark matter measurement\xspace}
\newcommand{\geant}{\mbox{\textsc{Geant4}}\xspace}

\newcommand{\fig}[1]{Figure~\ref{#1}\xspace}
\newcommand{\figs}[2]{Figures~\ref{#1} and \ref{#2}\xspace}
\newcommand{\tab}[1]{Table~\ref{#1}\xspace}
\newcommand{\sect}[1]{Sec.~\ref{#1}\xspace}

\journalname{Eur. Phys. J. C}

\begin{document}

\title{
Simulation and background characterisation of the SABRE South experiment
}

\subtitle{The SABRE South Collaboration}

\author{E.~Barberio\thanksref{melbourne,arc}
  \and
  T.~Baroncelli\thanksref{melbourne,arc}
  \and
  L.~J.~Bignell\thanksref{anu,arc}
  \and
  I.~Bolognino\thanksref{adelaide,arc}
  \and
  G.~Brooks\thanksref{swinburneng,arc}
  \and
  F.~Dastgiri\thanksref{anu,arc}
  \and
  G.~D'Imperio\thanksref{roma}
  \and
  A.~Di~Giacinto\thanksref{lngs}
  \and
  A.~R.~Duffy\thanksref{swinburne,arc}
  \and
  M.~Froehlich\thanksref{anu,arc}
  \and
  G.~Fu\thanksref{melbourne,arc}
  \and
  M.~S.~M.~Gerathy\thanksref{melbourne,arc}
  \and
  G.~C.~Hill\thanksref{adelaide,arc}
  \and
  S.~Krishnan\thanksref{swinburneng,arc}
  \and
  G.~J.~Lane\thanksref{anu,arc}
  \and
  G.~Lawrence\thanksref{swinburne,arc}
  \and
  K.~T.~Leaver\thanksref{adelaide,arc}
  \and
  I.~Mahmood\thanksref{melbourne,arc}
  \and
  A.~Mariani\thanksref{princeton}
  \and
  P.~McGee\thanksref{adelaide,arc}
  \and
  L.~J.~McKie\thanksref{anu,arc}
  \and
  P.~C.~McNamara\thanksref{anu,arc,e2}
  \and
  M.~Mews\thanksref{melbourne,arc}
  \and
  W.~J.~D.~Melbourne\thanksref{melbourne,arc}
  \and
  G.~Milana\thanksref{swinburneng,arc}
  \and
  L.~J.~Milligan\thanksref{melbourne,arc}
  \and
  J.~Mould\thanksref{swinburne,arc}
  \and
  F.~Nuti\thanksref{melbourne,arc,e1}
  \and
  V.~Pettinacci\thanksref{roma}
  \and
  F.~Scutti\thanksref{swinburne,arc}
  \and
  Z.~Slavkovsk\'{a}\thanksref{anu,arc}
  \and
  N.~J.~Spinks\thanksref{anu,arc}
  \and
  O.~Stanley\thanksref{melbourne,arc}
  \and
  A.~E.~Stuchbery\thanksref{anu,arc}
  \and
  G.~N.~Taylor\thanksref{melbourne,arc}
  \and
  C.~Tomei\thanksref{roma}
  \and
  P.~Urquijo\thanksref{melbourne,arc}
  \and
  C.~Vignoli\thanksref{lngs}
  \and
  A.~G.~Williams\thanksref{adelaide,arc}
  \and
  Y.~Y.~Zhong\thanksref{anu,arc}
  \and
  M.~J.~Zurowski\thanksref{melbourne,arc,e3}
}

%If needed can add email to some of the authors including the tags below as one of the \thanksref arguments
\thankstext{e1}{e-mail: francesco.nuti@unimelb.edu.au (corresponding author)}
\thankstext{e2}{e-mail: peter.mcnamara@anu.edu.au (corresponding author)}
\thankstext{e3}{e-mail: madeleine.zurowski@utoronto.ca (corresponding author)}

\institute{School of Physics, The University of Melbourne, Melbourne, VIC 3010, Australia\label{melbourne}
          \and
          Department of Nuclear Physics and Accelerator Applications, The Australian National University, Canberra, ACT 2601, Australia\label{anu}
          \and
          Department of Physics, The University of Adelaide, Adelaide, SA 5005, Australia\label{adelaide}
          \and
          Centre for Astrophysics and Supercomputing, Swinburne University of Technology, Hawthorn, VIC 3122, Australia\label{swinburne}
          \and
          School of Engineering, Swinburne University of Technology, Hawthorn, VIC 3122, Australia\label{swinburneng}
          \and
          ARC Centre of Excellence for Dark Matter Particle Physics, Australia\label{arc}
          \and
          INFN - Sezione di Roma, Roma I-00185, Italy\label{roma}
          \and
          INFN - Laboratori Nazionali del Gran Sasso, Assergi (L’Aquila) I-67100, Italy\label{lngs}
          \and
          Physics Department, Princeton University, Princeton, NJ 08544, USA\label{princeton}
}

%\date{\today}
\date{Received: date / Accepted: date}

\maketitle

\begin{abstract}
SABRE (Sodium iodide with Active Background REjection) is a direct detection dark matter experiment based on arrays of radio-pure \naitl crystals. 
The experiment aims at achieving an ultra-low background rate and its primary goal is to confirm or refute the results from the \dama experiment. 
The SABRE \pop phase was carried out in 2020-2021 at the Gran Sasso National Laboratory (LNGS), in Italy. 
The next phase consists of two full-scale experiments: SABRE South at the Stawell Underground Physics Laboratory, in Australia, and SABRE North at LNGS. This paper focuses on SABRE South and presents a detailed simulation of the detector, which is used to characterise the background for dark matter searches including \dama-like modulation. 
We estimate an overall background of 0.72 \dru in the energy range 1--6~\kev primarily due to radioactive contamination in the crystals. Given this level of background and considering that the SA\-BRE South has a target mass of 50~kg, we expect to exclude (confirm) \dama modulation at $4~(5)\sigma$ within 2.5 years of data taking.
\end{abstract}

\section{Introduction}
\label{sec:intro}
Understanding the particle nature of dark matter~\cite{DMhistory,DMevidence1,DMevidence2} is one of the most important open problems in modern physics, with many concurrent search programs. One of the primary search techniques for dark matter is direct detection. The primary goal of this method is to measure the recoil energy released by the interaction between a dark matter particle and the detector target nuclei or electrons. Dark matter interactions are expected to oscillate throughout the year in the standard halo hypothesis for cold WIMPs~\cite{wimp}. Due to the relative velocity of the dark matter halo with respect to the Earth, which revolves around the Sun as it moves through the galaxy, we expect a sinusoidal trend with a maximum in June and minimum in December for dark matter candidates with mass below 200~GeV for \naitl targets~\cite{modulationth}.

Among the many direct detection experiments, \dama~\cite{DAMA2008} is the only one to have observed an annual mo\-du\-la\-tion signal compatible with dark matter. This experiment uses an array of $\sim$250 kg of \naitl crystals and has observed a modulation for almost two dec\-ades with a statistical significance of $12.9\,\sigma$~\cite{DAMAPhase1,DAMAPhase2}.
Despite its longevity and high significance, the \dama result is in contrast with the null observations from other direct detection experiments~\cite{xenon1t,LUX,SuperCDMS,XMASS}. These experiments use a different target material to \dama, and so the comparison of results require the assumption of specific dark matter interaction models or classes of models.

A model-independent test of the \dama modulation is therefore best achieved with an experiment that uses the same target material and detection technique. The two main ongoing \naitl experiments, COSINE~\cite{cosine} and ANAIS~\cite{anais}, have higher background levels than \dama and have not confirmed or ruled out the \dama modulation with a statistical significance of at least 3$\sigma$. Both experiments are expected to continue collecting data over the coming years. The COSINE collaboration is also working on the development of ultra-low background crystals with a background level below 1 \dru for the next generation of their experiment~\cite{cosine200}. COSINE’s most recent results find a modulation approximately half that of \dama’s consistent with both the modulation amplitude reported by \dama and the zero-modulation case~\cite{cosine}, motivating further study.

For this purpose, the Sodium iodide with Active Background REjection (SABRE) experiment~\cite{SABRE}, is designed to measure the dark matter annual modulation interaction rate with the achievement of an ultra-low background. 
Direct dark matter detectors are suitably well-shielded against external radiation and their background rate is driven by radioactive contaminants in the detector material and in the materials used for the construction of the ex\-pe\-ri\-men\-tal setup. Such radioactive contamination may come from long-lived, naturally occurring isotopes or from cosmogenic activation. Careful selection or development of radio-pure materials and equipment is therefore mandatory, as well as a detailed know\-ledge of residual radioactivity.
SABRE's \naitl
crystals, photosensors and all detector materials are designed to reach ultra-high radio-purity levels. In addition, a liquid scintillator veto allows for an active rejection of the residual background.

The SABRE \pop (SABRE PoP) phase has been carried out at the Gran Sasso National Laboratory (LNGS) with a single 3.4 kg \naitl crystal~\cite{nai33}. 
The next phase consists of two full-scale experiments: SABRE South at the Stawell Underground Physics Laboratory (SUPL), in Australia, and SABRE North at LNGS, in Italy.
SABRE South and North differ in their shielding designs. SABRE South will utilise a liquid scintillator system for in-situ evaluation of some of the crystal contaminants, background rejection and particle identification of external background. SABRE North has chosen to adopt a fully passive shielding design~\cite{Calaprice:2022dxb} as the use of organic scintillators has been indefinitely phased out by LNGS.

This paper focuses on the SABRE South detector: we provide a brief description of the experiment and present a model of the expected background based on a Monte Carlo simulation and measurements of material radiopurity.
The energy spectrum expected in the SABRE \naitl crystals due to radioactive background processes is determined. 
We focus on the 1--6~\kev energy range, which is the region of interest (ROI) to study the \dama modulation and provide a prediction of the discovery and exclusion power of the experiment to such signal. We also provide the expected total time-dependent background rate through the lifetime of the experiment.

\section{Detector design and implementation into simulation}
\label{sec:southsetup}

\subsection{Technical design} 
\label{tech_des}
\begin{figure}
\centering
  \subfigure[The SABRE South Experimental setup.]
   {\includegraphics[width=0.4\textwidth]{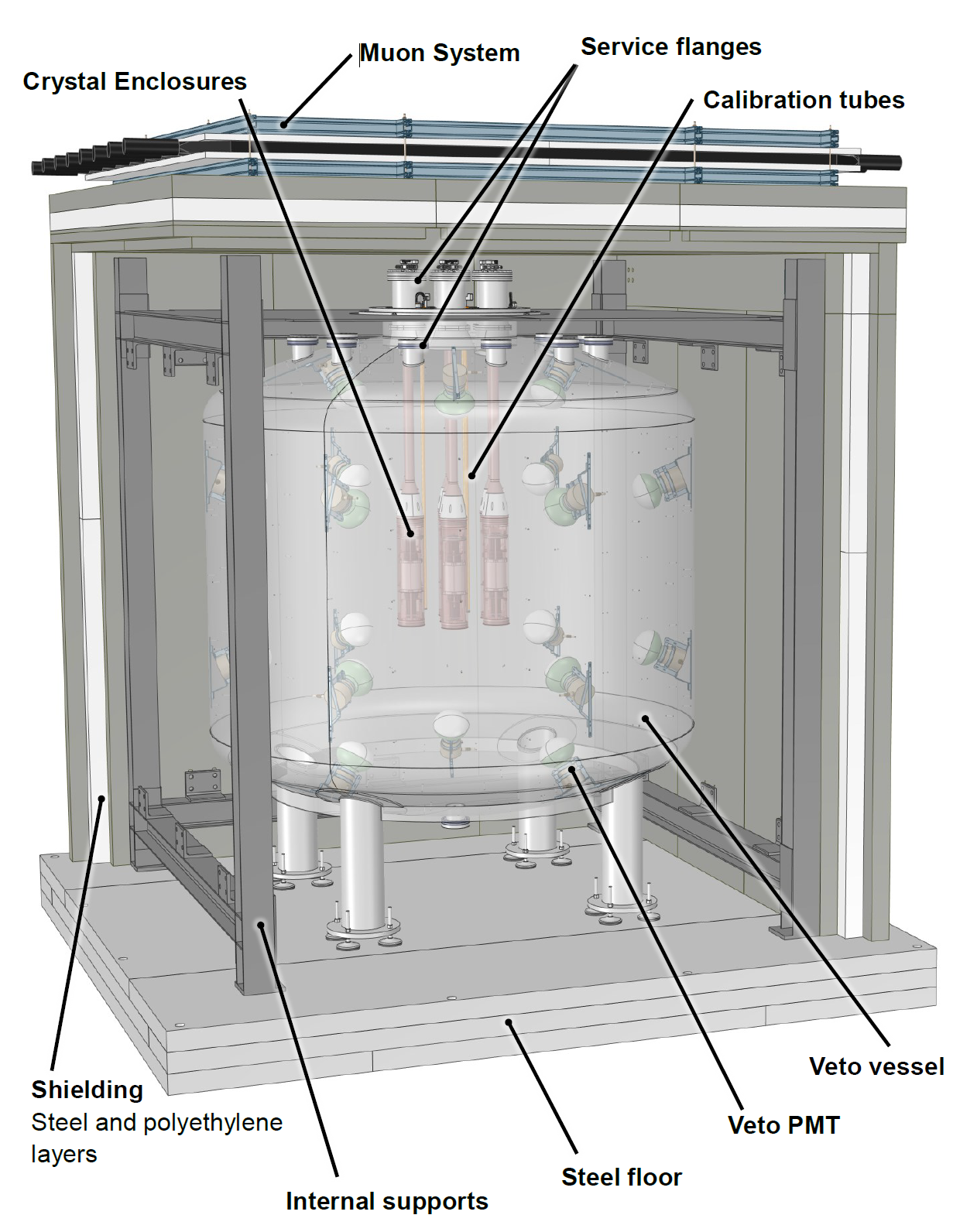}}
  \hspace{-5mm}
  \subfigure[Implementation of the experimental setup in the \geant simulation.]
   {\includegraphics[width=0.44\textwidth]{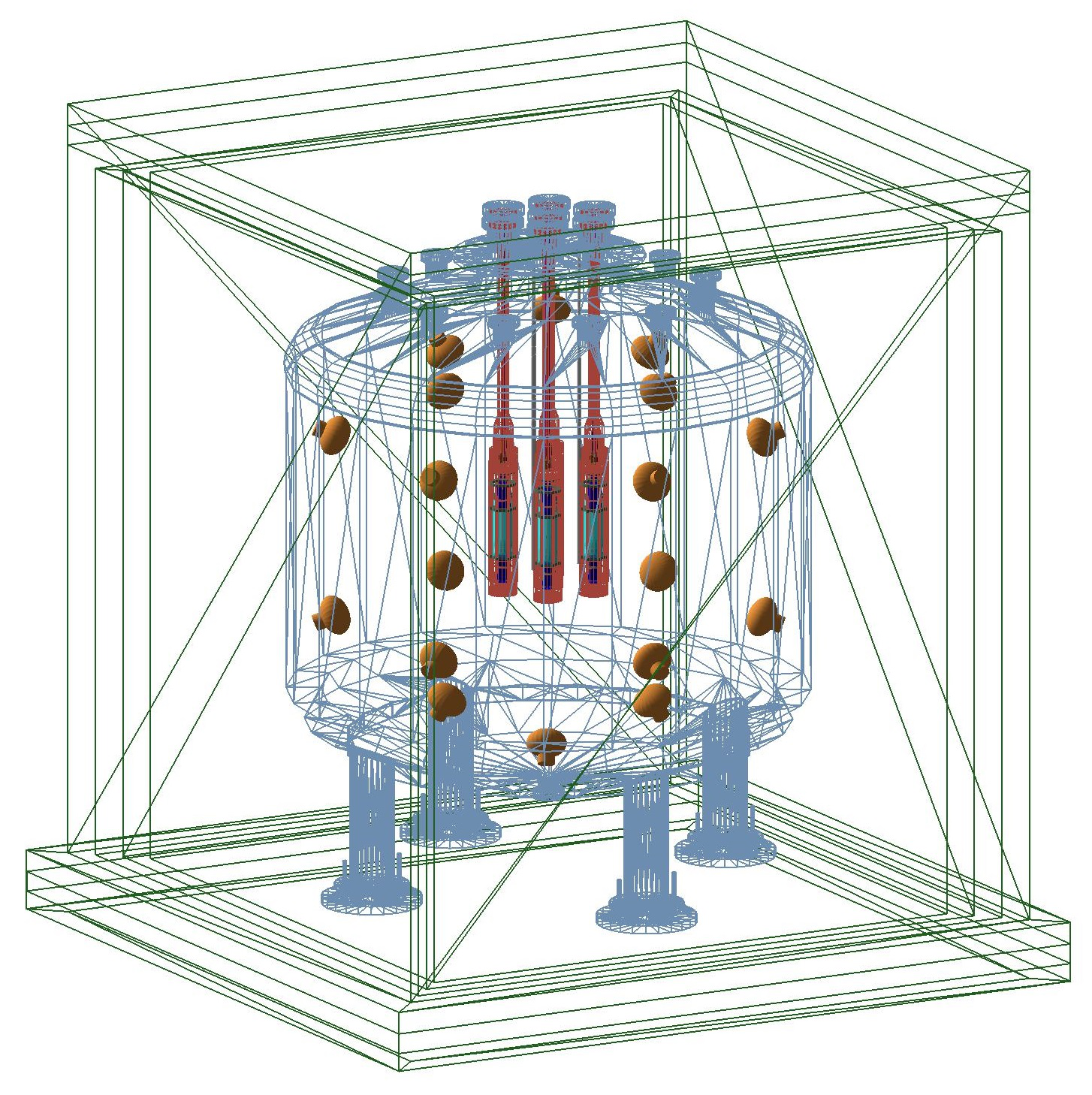}}
 \hspace{-5mm}
  \subfigure[Detailed view of a crystal detector module as modelled in \geant. The crystal is shown in cyan, PMTs in blue, PTFE supports in dark green, and copper parts in maroon. The external cylindrical copper enclosure is not shown.]
  {\includegraphics[width=0.46\textwidth]{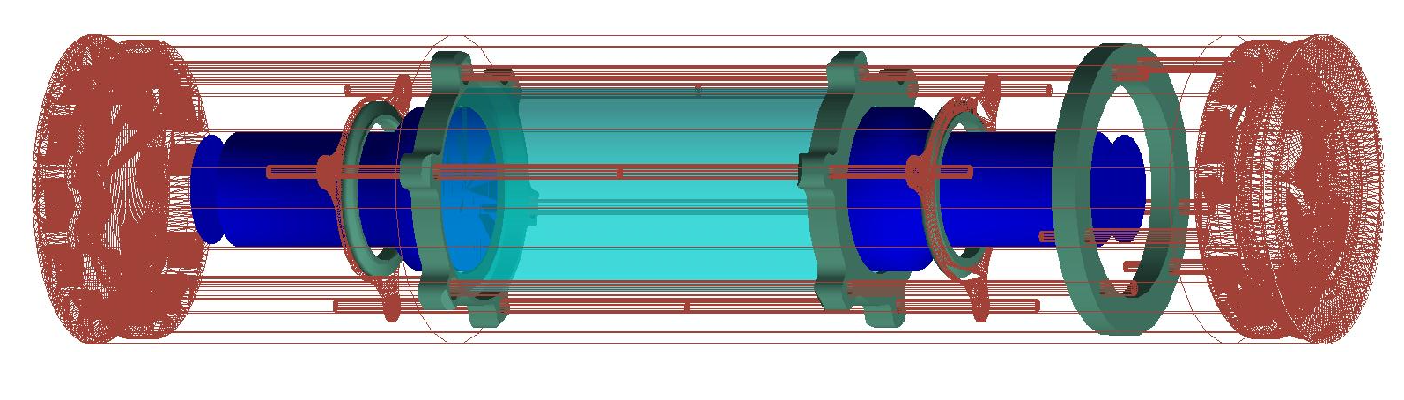}}
  \hspace{-5mm}
  \caption{The SABRE South experiment~(a), its representation in the simulation~(b) and a close-up view of a simulated crystal detector unit~(c).}
  \label{fig:SimulatedGeometry}
\end{figure}

SABRE South is made up of three different subdetector systems: the \naitl crystal detector system, the liquid scintillator veto system, and the muon paddle detectors. 
The crystals and the liquid scintillator veto system are further shielded by steel and polyethylene walls. 
The full experimental setup is shown in \fig{fig:SimulatedGeometry}(a).

The experiment can host seven \naitl cylindrical crystals 25 cm long and 5 cm in radius for a  mass of 7.2 kg per crystal (50.4 kg in total). Crystals will be grown from Merck's Astrograde powder, the highest purity NaI powder commercially available, which has a potassium contamination below 10~ppb, and uranium and thorium contamination below 1~ppt. These crystals are encapsulated in cylindrical oxygen-free high-thermal-conductivity (OFHC) copper enclosures flush\-ed with nitrogen.
These enclosures, shown in \fig{fig:SimulatedGeometry}(c), are composed of a hollow cylinder and two endcaps that seal the ends of the cylinder. The cylinder has a radius of 71.5 mm, a length of 664.5 mm and a thickness of 3 mm. The endcaps are up to 36 mm thick, but their surfaces are milled down to 5 mm thickness wherever possible to minimise the amount of material.
Each enclosure contains a number of different components. At the centre is a \naitl cylindrical crystal with a length of 25 cm and a radius of 5 cm, shown in \fig{fig:SimulatedGeometry}(c) in cyan. This is wrapped in PTFE foil, bookended by PTFE crystal holders, and coupled to a Hamamatsu R11065 PMT (76 mm diameter) on each side (shown in dark green and blue respectively). These components are all held together by 9 internal support rods made of copper. Three are 82.5 mm long and connect the top endcap 
to the inner PTFE ring. Three more rods run from this ring through the two PTFE crystal holders along the length of the crystal to the bottom endcap and are 553.5 mm long. Finally, three connect the two PMT holders, which also pass through the crystal holders and have a length of 399 mm. The enclosures are submerged in the veto vessel and held in place by copper conduits that also allow for cabling transport out of the vessel.

The SABRE South veto vessel is made of stainless steel (lined with Lumirror\texttrademark) approximately 3 m tall with a 2.6 m diameter at its widest, and is designed to hold 10 tonnes of liquid scintillator. The main body of the vessel is a cylinder of height 1.65 m and radius of 1.3 m and it is connected to spherical section endcaps. The interior of the vessel can be accessed from the top through a 70~cm diameter flange. Seven subflanges with a radius of 74.1 mm are mounted on top of the 70~cm diameter flange, to be used for crystal insertion. There is one subflange in the center and six surrounding it disposed at the vertices of a hexagon. The distance between the axes of any adjacent pair of subflanges and thus \naitl crystals is 26 cm. Twelve more small flanges with a radius of 50.75 mm are mounted on the top spherical endcap and are used for services, such as veto PMT cables and fluid handling. A triplet of aluminium pipes is placed equidistant from the crystal detector modules and allows the insertion of radioactive sources for calibration purposes.

The vessel is filled to a height of 2.42 m with liquid scintillator and topped  off with a nitrogen blanket. The scintillator itself is a mixture of Linear Alkyl Benzene (LAB) and the fluorophores PPO and Bis-MSB. The vessel is instrumented with eighteen Hamamatsu R5912 PMTs (204 mm diameter) to detect signals in the liquid scintillator. 
The PMTs are arranged in four horizontal planes: in the two central planes there are six PMTs equally spaced around the perimeter, while the upper and lower planes have just three equally spaced PMTs.
This detector system is able to observe around 0.12 photoelectrons/\kev, giving a detection threshold of around 50~\kev, and reaching 100\% efficiency at 200~\kev and above.

The vessel is then placed within a shielding system which has a total thickness of 26~cm on the top, bottom and sides. This is made up of a 10~cm thick layer of high-density poly\-ethylene (HDPE) to shield from neutrons, which is sandwiched between two 8~cm layers of low-carbon aluminium killed steel to shield from high-energy gamma rays. This structure has a cuboid shape and is around 3.5~m tall and 2.9~m wide for a total mass of almost 100 tonnes. The eight EJ200 muon detector paddles sit atop this shielding forming one continuous layer covering an area of 9.6~m$^2$ centred above the crystals. Each paddle is instrumented with two Hamamatsu R13089 PMTs (51 mm diameter). These have a timing resolution of 200~ps, allowing for position reconstruction along the length of the paddle within 5~cm. The energy threshold of the muon detector is approximately 1~\mev and muons vertically crossing the plastic scintillator are expected to deposit about 10~\mev of energy. 

This full setup will be located at the Stawell Underground Physics Laboratory (SUPL), 1025 m underground in Victoria, Australia, providing a 2900 m water-equivalent flat overburden. Assembly of the SABRE South detector at SUPL is commencing in 2023. SABRE South is expected to be ready to take data with a full set of crystals by 2025.

\subsection{\geant implementation}
We propagate radiation through the detector with the \geant simulation toolkit, version 10.7 \cite{geant42003,geant42006}. The physical characteristics of the experiment are reproduced in great detail in the simulation, as shown in \fig{fig:SimulatedGeometry}(b).
The experimental components with the characteristics described in \sect{tech_des} are replicated in the simulation with the exception of the muon detector, as it is made of low-radioactivity plastic and any residual radioactivity is attenuated by the shielding and so contributes negligibly to the total background model. For this reason, 
the simulation is more detailed closer to the crystal because the majority of the background rate comes from the crystal array itself.

The simulation accurately reproduce components' thicknesses and shapes as these affect both radiation emission rates from the materials and the absorption probability in the materials.
For the most complex component of the detector enclosure, the CAD design was transformed into gdml format and imported into the simulation. The crystal PMTs are modelled as a quartz window (with a 38 mm radius), a ceramic photocathode, a feedthrough plate, a PTFE voltage divider and a Kovar body. The vessel PMTs are modelled as an ellipsoid window with a radius of 9.5 cm and body made of borosilicate.
Outside of the crystal enclosures, the vessel has all the main features of the real one but omits small items such as stud bolts, cabling, nuts and fine machining. The overall size, thicknesses and mass match those of the real vessel. The simulated shielding does not include the internal steel support frame, which amounts to approximately $10\%$ of the total shielding mass. Components external to the enclosures lead to a background that is at least two orders of magnitude smaller than the crystal modules and therefore further precision was not considered necessary. 

For the SABRE simulation, we have chosen the shielding physics list recommended for underground low-back\-ground experiments, with the addition of the \geant ``option 4" for the electromagnetic (EM) physics~\cite{geant42016}. 
The package for EM interactions includes the Wentzel VI model at high energy, Msc95 model below 100 MeV~\cite{wentzel}, photon models from Livermore and Penelope, and Livermore ionisation model for electrons \cite{Livermore, Penelope}. 
The hadronic interaction model includes elastic, inelastic, capture and fission processes; precision models are used for neutrons with energy below 20 MeV.
The production and transport of optical photons both in crystal and in the LS veto have not been included in the simulation results described here; however, their inclusion is being pursued currently. 

\section{Radioactive contamination in the detector}
\label{sec:contaminations}
Radioactive decays form the vast majority of background for dark matter detection with this apparatus. The most relevant sources of radioactive contamination are primordial radionuclides ($^{40}$K and isotopes from the $^{238}$U and $^{232}$Th chains), anthropogenic radionuclides (e.g.$^{137}$Cs), cosmo\-geni\-cally-activated radionuclides (e.g. $^{3}$H) and environmental radioactive noble gases, such as $^{222}$Rn and $^{220}$Rn.
Contamination in the detector materials, especially in the \naitl crystal and the surrounding components, is responsible for nearly all of the background rate. Therefore, a thorough assessment of the level of radioactive contamination for every component of the experimental setup is needed.
The contamination levels of the materials composing the SABRE South experiment are based on screening techniques such as gamma ray spectrometry using High-Purity Ge (HPGe) detectors, neutron activation analysis (NAA), Accelerator Mass Spectrometry (AMS), and Inductively Coupled Plasma Mass Spectrometry (ICP-MS).

In the following sections we list the radioactive contamination of materials used for background calculation. In some cases where the contamination level is below the sensitivity of the measurement, we consider the latter as an upper limit and conservatively use it in the simulation. Secular equilibrium in the U and Th decay chains is assumed, unless otherwise specified.

A summary of the experimental components considered in the simulation as sources of radiation, with the corresponding materials and masses, is reported in \tab{tbl_masses}. 

\begin{table}[htbp!]
\footnotesize
\centering
\begin{tabular}{llll}
\hline
\bf Volume Name & Material & \# & Mass [kg]\\
\hline
\bf Crystal & \naitl & 7 & 50.4 \\
 \hline
\bf Crystal Wrapping  &  PTFE & 7 &  $2.6 \cdot 10^{-1}$\\
 \hline
 \multicolumn{4}{l}{{\bf Crystal Enclosure}} \\
Outer body &  Copper & 7 &  $51.6$ \\
Inner structure &  Copper & 7 &  $37.8 $ \\
Inner structure &  PTFE & 7 &  $2.8$ \\
 \hline
 \multicolumn{4}{l}{{\bf Crystal PMTs}} \\
Window & Quartz & 14 & $ 4.2 \cdot 10^{-1}$ \\
Body & Kovar & 14 & $ 1.3 $ \\
Feedthrough Plate & Ceramic & 14 & $ 2.2 \cdot 10^{-1}$ \\
 \hline
 \multicolumn{4}{l}{{\bf Veto}} \\
 Scintillator & LAB & 1 & $ 1.0 \cdot 10^{4} $ \\
 Vessel & Stainless steel & 1 & $ 1.8 \cdot 10^{3} $ \\
 PMTs & Borosilicate glass & 18 & 8.9\\
\hline
 \multicolumn{4}{l}{{\bf Services}} \\
 Crystal conduits & Copper & 7 & $54.4$ \\
 Calibration pipes & Aluminium & 3 & $4.7$\\
 \hline
 \multicolumn{4}{l}{{\bf Shielding }}\\
 Polyethylene layer & HDPE & 1 & $6.6 \cdot 10^{3} $ \\
 Steel layers & Carbon Steel & 2 & $9.0 \cdot 10^{4} $ \\
\hline
\end{tabular}
  \caption{Components of the SABRE South experiment that have been implemented in the \geant simulations, as well as their component materials, the number of times the components are used and their total mass.}
  \label{tbl_masses}
\end{table}

\subsection{\naitl Crystals}
\label{sec:cont_crystal}
Contamination sources within the \naitl crystals can be split into two categories: radiation due to naturally occurring radioisotopes in \naitl powder (radiogenic contamination), and radiation due to activation of the Na and I when exposed to high-energy cosmic rays (cosmogenic contamination). 
Radiogenic contamination is composed of ultra-long-lived isotopes, producing a background rate that is nearly constant over the lifetime of the detector. The fastest decaying components of the background are from $^{85}$Kr and from $^{210}$Pb and daughters, which we allow to not be in secular equilibrium with $^{238}$U. Their contributions are expected to be reduced after five years by 27\% and 14\%, respectively.
Cosmogenic activation is significant above ground, but irrelevant when the crystal is placed hundreds of meters underground, since the cosmic ray flux and thus the activation rate is millions of times lower. Thus, the concentration of cosmogenic isotopes in the crystals is strongly dependent on the travel time and route from the site of crystal growth to SUPL, and the extent of `cool-down' time allowed in the underground laboratory prior to the start of data taking. As a general rule, the cosmogenic isotopes have much shorter half-lives compared to radiogenic isotopes, leading to a time-dependent decaying background signature in the detector. The contamination levels of each are discussed separately in this section. For simplicity, we have assumed the same background model for all seven crystals.

For radiogenic radioisotopes, we use the values measured on the \naitl crystal grown for the SABRE PoP~\cite{calaprice2021,nai33} where the activity of $^{40}$K, $^{210}$Pb, and $^{129}$I were determined from spectral analysis. In addition to this, $^{210}$Pb is further constrained by alpha counting, and $^{40}$K via ICP-MS measurements on crystal off-cuts. Upper limits for $^{87}$Rb, $^{238}$U and $^{232}$Th in the crystal powder are also measured with ICP-MS \cite{PoP_MC}. 

\begin{table}[htbp!]
\footnotesize
\centering
\begin{tabular}{cc}
\hline
Isotope & Activity [mBq/kg]                                                            \\  \hline
$^{40}$K                                       & $1.4 \cdot 10^{-1}$  \\
$^{238}$U                                     & $<5.9 \cdot 10^{-3}$                         \\
$^{232}$Th                                     &  $<1.6 \cdot 10^{-3}$  \\
$^{87}$Rb                                      & $<3.1 \cdot 10^{-1}$                                                          \\
$^{210}$Pb                                    &  $4.1 \cdot 10^{-1}$ \\
$^{85}$Kr                                      & $<1.0 \cdot 10^{-2}$                                                           \\
$^{129}$I                                      &  1.3                \\ \hline
\end{tabular}
\caption{Activity levels of radiogenic isotopes in the SABRE \naitl crystals. 
Values for $^{40}$K, $^{210}$Pb, $^{85}$Kr and $^{129}$I were measured from the crystals~\cite{calaprice2021,nai33}, while the other isotopes from the powder~\cite{PoP_MC}.
}
\label{tab:intrinsic-rad}
\end{table}

Cosmogenically-induced contamination in the crystals is also critical for dark matter searches. The exposure of \naitl to cosmic rays at sea-level leads to the production of thousands of radioisotopes per kg of crystal per hour. It is therefore necessary to minimise the time that the crystal spends above ground. 
We expect that the production of a \naitl crystal at RMD (Radiation Monitoring Devices), Massachu\-setts, USA will take up to two months and the transportation by sea to SUPL, Victoria, Australia will take another month. Transportation via plane has been considered but disregarded due to higher cosmogenic activation despite the shorter transit time. Cosmogenic activation can also occur in the \nai powder prior to crystal growth. In this paper, we assume that the activation in the \nai powder is negligible compared to the activation in the \naitl crystal, since radio impurity levels are mitigated by the process of powder preparation and crystal growth. 

Among the cosmogenic radioisotopes, $^3$H is especially problematic as it is long-lived ($T_{1/2} = 12.3$yr) and has a continuous beta decay spectrum with a 18.6~keV Q-value, leading to a background in the low-energy region of interest for dark matter detection. The COSINE-$100$ experiment has shown that the $^3$H activity in its \naitl crystals grows at a rate of about 0.18 mBq/kg per year of surface exposure~\cite{cosinebg2020}. Thirteen other key cosmogenic isotopes have also been selected based on a cosmic-like neutron irradiation experiment and a \geant neutron activation simulation.

The isotopes $^{109}$Cd, $^{113}$Sn, $^{121}$Te and $^{126}$I can also deposit electrons at the low-energy range of interest to dark matter experiments. $^{126}$I has a half-life of only $12.9$ days, and so should be absent inside the NaI(Tl) just months after underground placement, while the other three radioisotopes will persist for a few years. $^{121}$Te is as short-lived as $^{126}$I, however it is continuously regenerated by the decay of its long-lived parent $^{121m}$Te. An equilibrium is reached quickly as $^{121}$Te decays at the same rate as $^{121m}$Te.
Similarly, equilibrium is developed between $^{109}$Cd, $^{113}$Sn, $^{127m}$Te and their corresponding radio-daughters $^{109m}$Ag, $^{113m}$In and $^{127}$Te. 

The cosmogenic activation of the SABRE South \naitl crystals is calculated with the ACTIVIA~\cite{ACTIVIA} simulation software package. Activation during manufacturing and transportation to SUPL is corrected to take into account altitude, geomagnetic shielding and solar activity. The attenuation of the cosmogenic flux due to crystals being grown indoors and shipped in a cargo ship is not considered, therefore this estimate is expected to be slightly conservative. Measurements of activation on \naitl crystals show that ACTIVIA's predictions are accurate for some isotopes, but are up to a factor 6 different for other isotopes~\cite{cosinebg2020}. In particular, the activation rate of $^3$H is measured 2.3 times higher than what expected from calculation in multiple experiments~\cite{Amar__2018,PhysRevD.107.022006}. We use this factor to correct the prediction of $^3$H activity obtained from ACTIVIA. The activation rates of $^{109}$Cd, $^{113}$Sn, $^{121m}$Te and $^{127m}$Te appear to be overpredicted by ACTIVIA. We did not scale down these activities to keep our background estimate conservative.

We consider a six-month cool-down period, beginning when the crystals arrive underground at SUPL. During this period the activities of cosmogenically-induced isotopes decrease, especially for short-lived isotopes. 
The expected level of activity after this cool-down period is reported in \tab{tblbulk_NaI}, and has been used as input to our simulations. After several years of underground operation, the only cosmogenic isotope to remain particularly active would be $^3$H.

\begin{table}[htbp!]
\footnotesize
\centering
\begin{tabular}{ccc}
\hline
Isotope      &     Activity [mBq/kg]&  Half life [days]  \\ \hline
$^{3}$H      &     $9.4 \cdot 10^{-3}$   &   4496.8  \\
$^{22}$Na    &     $4.3 \cdot 10^{-2}$   &   949.7  \\
$^{109}$Cd   &     $5.3 \cdot 10^{-3}$   &   461.4  \\
$^{109m}$Ag  &     $5.3 \cdot 10^{-3}$   &   4.6 10$^{-4}$  \\
$^{113}$Sn   &     $1.44 \cdot 10^{-2}$  &   115.1  \\
$^{113m}$In  &     $1.41 \cdot 10^{-2}$  &   0.07  \\
$^{121m}$Te  &     0.16            &   164.2  \\
$^{121}$Te   &     0.16   &   19.2  \\
$^{123m}$Te  &     $8.35 \cdot 10^{-2}$  &   119.2  \\
$^{125m}$Te  &     $5.96 \cdot 10^{-2}$  &   57.4  \\
$^{127m}$Te  &     0.14            &   106.1  \\
$^{127}$Te   &     0.14            &   0.39  \\
$^{125}$I    &     0.19            &   59.4 \\
$^{126}$I    &     $1.0 \cdot 10^{-4}$   &   12.9  \\ 
\hline
\end{tabular} 
\caption{Radioactivity levels of cosmogenically-activated isotopes in the SABRE \naitl crystals calculated using ACTIVIA~\cite{ACTIVIA} after a 6 month cool-down period. The activity of the short-lived daughters $^{113m}$In, $^{121}$Te, $^{127}$Te, and $^{109m}$Ag are computed assuming equilibrium with their long-lived mothers $^{113}$Sn, $^{121m}$Te, $^{127m}$Te, and $^{109}$Cd, where their branching ratios are accounted for. \label{tblbulk_NaI}}
\end{table}

\subsection{Crystal PMTs and reflector foil}
\label{sec:cont_PMT}
Low-radioactive Hamamatsu R11065 PMTs are used to detect signals in the \naitl crystals. Extensive HPGe screening has been performed on Hamamatsu\\ R11410 PMTs~\cite{XenonPMTs}, which are identical to the R11065 model except for dimensions and for the photocathode material.
%The radioactivity levels of the crystal PMTs are based on measurements by the XENON Collaboration~\cite{XenonPMTs}. They have performed extensive HPGe screening of the Hamamatsu\\ R11410 PMTs. This PMT is identical to the R11065 model used by SABRE South except for dimensions and for the photocathode material. 
The measurement shows that the highest contributions in terms of radioactivity come from the Kovar body, the quartz window and the ceramic feed\-thro\-ugh plates. No significant contribution was attributed to the tiny amount of photocathode material; thus we assume the same for the R11065 model.

In our simulations, we model the crystal PMTs as the assembly of the three higher-radioactivity components, namely the Kovar body, the quartz window and the ceramic feed\-through plates. The contamination values assigned to them are reported in \tab{bulk_PMT} and have been calculated from the values measured on R11410 PMTs~\cite{XenonPMTs}, and adjusted for the difference in size/mass of each component. 
We also scale up each activity for a correction factor to account for the higher radioactivity levels of some isotopes, such as $^{40}$K and $^{60}$Co, found in the assembled R11410 PMT compared to the sum of the single parts. The summed activities from the three components match the total measured value from Table 5 of Ref.~\cite{XenonPMTs} and at the same time the ratios of activity levels in the three parts are kept constant and equal to those measured in the raw materials.
No contamination from $^{235}$U and $^{137}$Cs was detected, 
thus they have not been considered in this work.

\begin{table}[htb!]
\footnotesize
\centering
\begin{tabular}{cccc}
\hline
&  \multicolumn{3}{c}{Activity [mBq/PMT]} \\
\cline{2-4}
\multicolumn{1}{c}{Isotope}  & Body & Window & Ceramic plate \\
\hline
$^{40}$K      & $<$5.9	      &   $<$0.48	& 6.5	 \\ 
$^{60}$Co     & 0.65	      &   $<$0.042	& $<$0.19	 \\ 
$^{238}$U     & $<$0.52	      &   $<$1.8	& 13	 \\
$^{226}$Ra    & $<$0.29	      &   0.040	        & 0.29	 \\
$^{232}$Th    & $<$0.0098     &   $<$0.037      & 0.70	 \\ 
$^{228}$Th    & $<$0.41	      &   $<$0.015      & 0.13	 \\ 
\hline 
\end{tabular}
\caption{Radioactivity levels of crystal R11065 PMT components, obtained from the values of R11410 PMTs components~\cite{XenonPMTs} rescaled to account for mass differences and total PMT radioactivity corrections. 
\label{bulk_PMT}}
\end{table}

PMTs are coupled to the crystal with Dow Corning optical silicone grease. Radiation originating from the optical grease and the PMT window are expected to have the same detection efficiency as these elements are positioned only a few $\mu$m apart. The Dow Corning radioactivity levels ($<1$~ppb $^{238}$U, $<1$~ppb $^{232}$Th, $1$~ppm $^{40}$K )~\cite{radiopurity} are five times lower than those of the PMT window for $^{238}$U, three times lower for $^{232}$Th and 27 times lower for $^{40}$K.
Moreover, the mass of the PMT window is greater by about 60 times than the mass of optical grease that will likely be used for the coupling. As such, the contribution of the optical grease to the radioactivity background is ne\-gli\-gi\-ble.

The PTFE reflector wrapped around the crystal can constitute a non-negligible source of background. Surface $^{210}$Pb contamination was measured during the background characterization studies of the first SABRE \naitl crystal detector~\cite{calaprice2021}. We are looking to source PTFE reflector with better radiopurity. One option is to obtain the same PTFE tape which is used by CUO\-RE-0, and which has a much smaller $^{210}$Pb surface contamination~\cite{CUORE2NU}. We will test it and will also test PTFE reflectors from other suppliers. For this simulation we use the surface 210Pb contamination from CUORE-0~\cite{CUORE2NU}. 
We have not detect contamination of $^{238}$U, $^{232}$Th and $^{40}$K in the reflector from direct counting, so here we use the values measured by XENON with ICP-MS~\cite{XENONScreening}, which are typical for PTFE materials.
The PTFE reflector contamination values used in the simulation are listed in \tab{bulk_PTFEWrapping}.
\begin{table}[htbp!]
\footnotesize
\centering
\begin{tabular}{cc}
\hline
Isotope & Activity \\
\hline
$^{40}$K & 3.1 [mBq/kg] \\
$^{238}$U & 0.25 [mBq/kg] \\
$^{232}$Th & 0.5 [mBq/kg] \\
$^{210}$Pb & $3 \cdot  10^{-5}$ [mBq/cm$^{2}$] \\
\hline
\end{tabular}
\caption{Radioactivity levels of PTFE reflector foil~\cite{XENONScreening,CUORE2NU}. \label{bulk_PTFEWrapping}}
\end{table}

\subsection{Copper and PTFE parts}
\label{sec:cont_enclosure}
The crystal enclosures and the wiring conduits that connect them to the main body of the experimental vessel consist primarily of oxygen-free high-thermal-conductivity (OFHC) copper. Following from earlier work on the SABRE PoP~\cite{PoP_MC}, we use the same radionuclides and their levels of activity in this work, as the material is expected to have a similar radioactive contamination. This includes the $^{238}$U and $^{232}$Th decay chains as well as $^{40}$K for the radiogenic background~\cite{CUORE2NU}. We also consider cosmogenic activation in the copper and we use the saturation levels of the copper cosmogenic isotopes at sea level~\cite{Baudis2015}. This leads to a conservative estimate of the background from copper, as many dominant contributions come from short-lived $^{58}$Co and $^{56}$Co, which will eventually cool down once the component are brought underground.
%, which are based on levels measured by the CUORE collaboration \cite{CUORE2NU}. 
%Cosmogenically activated isotopes are also considered which were originally obtained from a study made by the XENON experiment \cite{Baudis2015}. 
The isotopes considered in the simulation for this material are listed in \tab{cont_copper} along with their expected activity levels.

\begin{table}[htb!]
\footnotesize
\centering

\begin{tabular}{ccc}
\hline
\multicolumn{3}{c}{Radiogenic} \\
\hline
	Isotope & \multicolumn{2}{c}{Activity [mBq/kg]} \\
\hline
$^{40}$K &   \multicolumn{2}{c}{0.7 } \\
$^{238}$U &   \multicolumn{2}{c}{0.065 }  \\
$^{232}$Th &   \multicolumn{2}{c}{0.002 } \\
\hline
\multicolumn{3}{c}{Cosmogenic} \\
\hline
  Isotope & Activity [mBq/kg] & Half life [days] \\
\hline
$^{60}$Co & 0.340 & 1925  \\
$^{58}$Co & 0.798 & 71  \\
$^{57}$Co & 0.519 & 272  \\
$^{56}$Co & 0.108 & 77  \\ 
$^{54}$Mn & 0.154 & 312  \\
$^{46}$Sc & 0.027 & 84  \\
$^{59}$Fe & 0.047 & 44  \\
$^{48}$V & 0.039 & 16  \\
\hline
\end{tabular}
\caption{Relevant isotopes and their radioactive activity levels assumed for the (OFHC) copper sections of the SABRE South experiment \cite{PoP_MC, Baudis2015, CUORE2NU}. \label{cont_copper}}
\end{table}

In addition to the copper parts, the crystal enclosure also includes some PTFE sections which consist of a ring to interface the upper and lower copper rods, two ring-like structures to hold the crystal in place and two more rings to help secure the PMTs. These sections are shown as the dark green sections in \fig{fig:SimulatedGeometry}(b). 
PTFE has typically excellent radiopurity and usually only upper limits can be set on isotopic activity. We use the upper limits measured with a HPGe detector by the XENON collaboration~\cite{XENONScreening}, shown in \tab{bulk_PTFE}, as we plan to measure our PTFE components with a detector of similar sensitivity.

\begin{table}[htb!]
\footnotesize
\centering
\begin{tabular}{cc}
\hline
Isotope & Activity [mBq/kg] \\
\hline
$^{40}$K &  $<$2.25   \\
$^{238}$U & $<$0.31   \\
$^{232}$Th & $<$0.16 \\
$^{60}$Co & $<$0.11  \\
$^{137}$Cs & $<$0.13 \\
\hline
\end{tabular}
\caption{Relevant isotopes and their radioactive activity levels 
of the PTFE sections which are part of the SABRE South crystal enclosures~\cite{PoP_MC,XENONScreening}. 
\label{bulk_PTFE}}
\end{table}

\subsection{Veto components: Stainless steel, PMTs and Liquid Scintillator}
\label{sec:cont_veto}
The veto detector contains 10 tonnes of liquid scintillator consisting of a linear-alkylbenzene (LAB) solvent, 3.5 g/L 2,5-diphenyloxazole (PPO) and 15 mg/L 1,4-bis-methylstyryl benzene (bisMSB). The purified LAB was supplied by Sino\-pec Jinling Petrochemical Co. Ltd, which is contracted to supply identical material for the JUNO experiment. 
Thus, we use the radioactive contamination levels measured by JUNO~\cite{JUNO_scint}. We additionally include a $^{7}$Be contamination at the level found by Borexino~\cite{BorexinoLS}, but this results in negligible background compared to the other isotopes. 
The activities of these radioisotopes are listed in \tab{bulk_ls}.
\begin{table}[htb!]
\footnotesize
\centering
\begin{tabular}{cc}
\hline
Isotope & Activity [mBq/kg] \\
\hline
$^{40}$K &  $2.71 \cdot 10^{-5}$  \\
$^{238}$U & $1.24 \cdot 10^{-5}$ \\
$^{232}$Th & $4.05 \cdot 10^{-6}$  \\
$^{210}$Pb & $2.08 \cdot 10^{-3}$   \\
$^{210}$Bi & $2.08 \cdot 10^{-3}$   \\
$^{7}$Be & $1.20 \cdot 10^{-6}$   \\
%$^{14}$C & $4.10 \cdot 10^{-1}$   \\ %Borexino
$^{14}$C & 1.65 \\ %JUNO
$^{39}$Ar & $5.81 \cdot 10^{-6}$   \\
$^{85}$Kr & $5.81 \cdot 10^{-6}$   \\
\hline
\end{tabular}
\caption{Radioactivity levels of the SABRE South LAB scintillator~\cite{JUNO_scint,BorexinoLS}. 
\label{bulk_ls}}
\end{table}

%Veto PMTs
The eighteen Hamamatsu R5912 PMTs used in the veto are primarily made from low-radioactivity borosilicate glass. The contamination of this PMT model have been measured by the DarkSide-50 collaboration~\cite{DS50PMT} and are reported in \tab{bulk_vetopmt}. 

\begin{table}[htbp!]
\footnotesize
\centering
\begin{tabular}{cc}
\hline
Isotope & Activity [mBq/PMT] \\
\hline
$^{40}$K & 649\\ 
$^{238}$U & 883\\
$^{232}$Th & 110\\
$^{235}$U & 41 \\
\hline
\end{tabular}
\caption{Radioactivity levels of the Hamamatsu R5912 Veto PMTs. Values are taken from \cite{DS50PMT}. \label{bulk_vetopmt}}
\end{table}

%Steel
The SABRE veto vessel is fabricated from low-radio\-activity stainless steel sourced from NIRONIT.
%Edelstahl GmbH \& Co. KG. 
The vessel construction was carried out by Tasweld and used unthoriated welding to minimise the radioactivity introduced during the fabrication. Samples of the stainless steel sheets used in the construction of the vessel were tested for radioactivity at LNGS using HPGe detectors. Quantities of $^{238}$U, $^{232}$Th and $^{60}$Co were measured at the level of mBq/kg. Contamination from $^{40}$K, $^{137}$Cs and $^{235}$U were not found, and $90\%$ confidence level upper limits were set. The averages of the measured radioactivity level or upper limit for each isotope are reported in \tab{bulk_steel}.
\begin{table}[htbp!] 
\footnotesize
\centering
\begin{tabular}{cc}
\hline
Isotope & Activity [mBq/kg] \\
\hline
$^{238}$U & 1.5\\
$^{232}$Th & 2.0\\
$^{235}$U & $< 1$\\
$^{40}$K & $< 8$\\ 
$^{60}$Co & 5\\ 
$^{137}$Cs & $< 0.5$\\ 
\hline
\end{tabular}
\caption{Averages radioactive levels measured on samples of the stainless steel used for the SABRE veto vessel. For $^{40}$K, $^{137}$Cs and $^{235}$U no contamination was found and the averages of the $90\%$ confidence level upper limits are provided instead. The measurements were performed at LNGS with HPGe detectors.\label{bulk_steel}}
\end{table}

\subsection{Passive Shielding}
\label{sect:ext_shield}

Samples of the Low Carbon Aluminium Killed steel forming the SABRE South passive shielding have been screened with a HPGe detector by the Australian Nuclear Science and Technology Organisation (ANSTO) in Australia. A sample of the SABRE South vessel steel was also measured for comparison. No radioactive isotope was detected in either sample. The measurement was approximately ten times less sensitive than that used for the vessel material screening (\tab{bulk_steel}). We consider conservative radioactive concentration limits in the steel based on these measurements. The radioactivity of the HDPE layers has not been measured since it is not expected to contribute significantly to the background,
so limits based on radioactive measurements from XENON~\cite{XENONScreening} are used for these background calculations.

\begin{table}[htbp!] 
\footnotesize
\centering
\begin{tabular}{ccc}
\hline
&  \multicolumn{2}{c}{Activity [mBq/kg]} \\
\cline{2-3}
Isotope & Steel & HDPE \\
\hline
$^{238}$U & $<$13 & 0.23\\
$^{232}$Th & $<$6.7 & $<$0.14\\
%$^{235}$U & - & $<$0.37\\
$^{40}$K & $<$110 & 0.7\\ 
$^{60}$Co & $<$5.5 & 0.06 \\ 
$^{137}$Cs & $<$6.0 & $<$1.4 \\ 
\hline
\end{tabular}
\caption{Radioactivity levels of the shielding steel and HDPE components. The steel radioactivity has been measured by ANSTO, while the HDPE values are from XENON~\cite{XENONScreening}.
\label{shielding_activity}}
\end{table}

\subsection{Muon detector}
\label{sect:muon_det}
The muon detector is located above the passive shielding and the radiation from its components is highly attenuated by the passive shielding and the active veto system. From approximate calculations, the contribution to the background rate in the crystals is expected to be negligible, so we have not pursued a dedicated simulation estimate. This  calculation is performed by assuming that the contamination in the plastic scintillator is at a similar level to the PTFE in  \tab{bulk_PTFE} and that the contamination in the PMTs is equivalent to the R5912 PMTs in \tab{bulk_vetopmt}. From the simulation of radiation emanated by the top outer layer of the passive shielding we obtain a conservative estimate of the probability of detecting radiation originating from the muon detector. This radiation would travel indeed through more material in order to reach the crystals than that produced in the shielding, and the additional distance of the muon detector from the crystals also reduces the acceptance. Based on these arguments, the contribution of the muon detector to the background is of the order of $10^{-12}$~\dru in the range 1--6~\kev, which is completely negligible.

\subsection{External radiation}
\label{sect:ext_rad}
Preliminary measurements of environmental radiation at the designated location for the SABRE South experiment were performed prior to the construction of SUPL with detectors provided by ANSTO. We use these measurements to assess the order of magnitude of the external radiation background. More accurate measurements will be performed once the laboratory is ready to host background detectors.

The gamma-ray energy spectrum was measured with a 3"$\times$3" \naitl detector. The integrated flux above 100~keV amounts to 2.5~cm$^{-2}$s$^{-1}$. We expect lower background flux after the completion of the laboratory as it has been constructed with materials that are less radioactive than the surrounding cave rock, providing passive shielding for  radioactivity originating from the rock.

Neutron flux was measured with two BF$_{3}$ proportional tubes, one of which was surrounded by a 15" diameter HDPE cylinder. The naked detector was used to measure thermal neutrons, while the shielded detector is sensitive to MeV energy neutrons. We have measured a thermal neutron flux of $3 \cdot 10^{-5}$~cm$^{-2}$s$^{-1}$ and a fast neutron flux of $7 \cdot 10^{-6}$~cm$^{-2}$s$^{-1}$.
The fast neutron energy spectrum was not measured but it was calculated using SOURCES~\cite{SOURCES}. We considered neutron production from ($\alpha$,n) transfer reactions and spontaneous fission in the rocks. Since the neutron energy spectra for these processes are very similar, for simplicity we use the spectrum of ($\alpha$,n) reactions from the $^{232}$Th decay chain, which is expected to be the predominant component. We did not consider the neutron spectrum from underground cosmogenic activation since this process is negligible at the depth of the experiment (2.9~km~w.~e.).

\section{Results}\label{sec:Results}
We simulated radioactive decays in different components of the setup, according to the radioactive contamination of the materials described in the previous sections.
For every combination of isotope and location inside the setup, we simulated a number of events large enough to keep the statistical uncertainty in the output spectrum well below the percent level for the crystal background and below a few percent for the outermost volumes.  
While the simulation of radioactive decays in the detector materials has been done within a single \geant simulation, the background due to external radiation has been simulated in steps. This is necessary because the probability of external radiation propagating through the detector is extremely small, making it prohibitive to produce enough statistics for a meaningful result. We have instead separately simulated the propagation though the passive shielding and through the vessel. We have also assumed that radiation originates uniformly around the detector, and is directed towards the crystal array. Since this latter assumption would significantly inflate the background induced in the crystals, correction factors based on the geometric acceptance of the system are applied.

The simulation records the energy deposited in the crystals and the liquid scintillator. The optical simulation, which includes the generation, propagation and collection of optical photons from scintillation, is not carried out in this study. Inclusion of optical photon propagation for each simulated radioactive event is computationally prohibitive.
However, dedicated simulations with optical physics were used to assess energy thresholds and resolutions of the detectors. We expected marginally better thresholds and resolutions for the liquid scintillator system compared to the SABRE PoP and crystal quality equivalent to that of NaI-33~\cite{nai33}. We apply an energy-dependent Gaussian smearing to the energies recorded in the simulation, using $\sigma(E) = 1.4 \% \cdot\sqrt{E} ~[\text{MeV}]$ for the crystal signal and $\sigma(E) = 15.4 \% \cdot\sqrt{E} ~[\text{MeV}]$ for the liquid scintillator. They correspond to a resolution of $5.73\%$ at 59.5~keV ~\cite{calaprice2021} and $9.5\%$ at 2615~keV~\cite{mariani}, respectively. The energy threshold is set to 1~\kev for the crystal detectors by design, and 50~\kev for the liquid scintillator based on optical simulations.

The background contributions to the measurement of single low-energy crystal interactions compatible with dark matter and the \dama annual modulation were then evaluated. We require a single crystal detector interaction with energy above threshold ($>1$~\kev) and no energy in the veto system above 50~\kev (the ``veto requirement''). %This will be referred to hereafter simply as the ``veto requirement''. 
We focus on low-energy crystal interactions in the range 1--6~\kev as this is where \dama observes the annual modulation and where the WIMP signal is expected to appear. In the following, we refer to this set of requirements as the \dmm region.

\subsection{SABRE South background model}
\begin{figure}[htb!] \centering
    \includegraphics[width=0.46\textwidth]{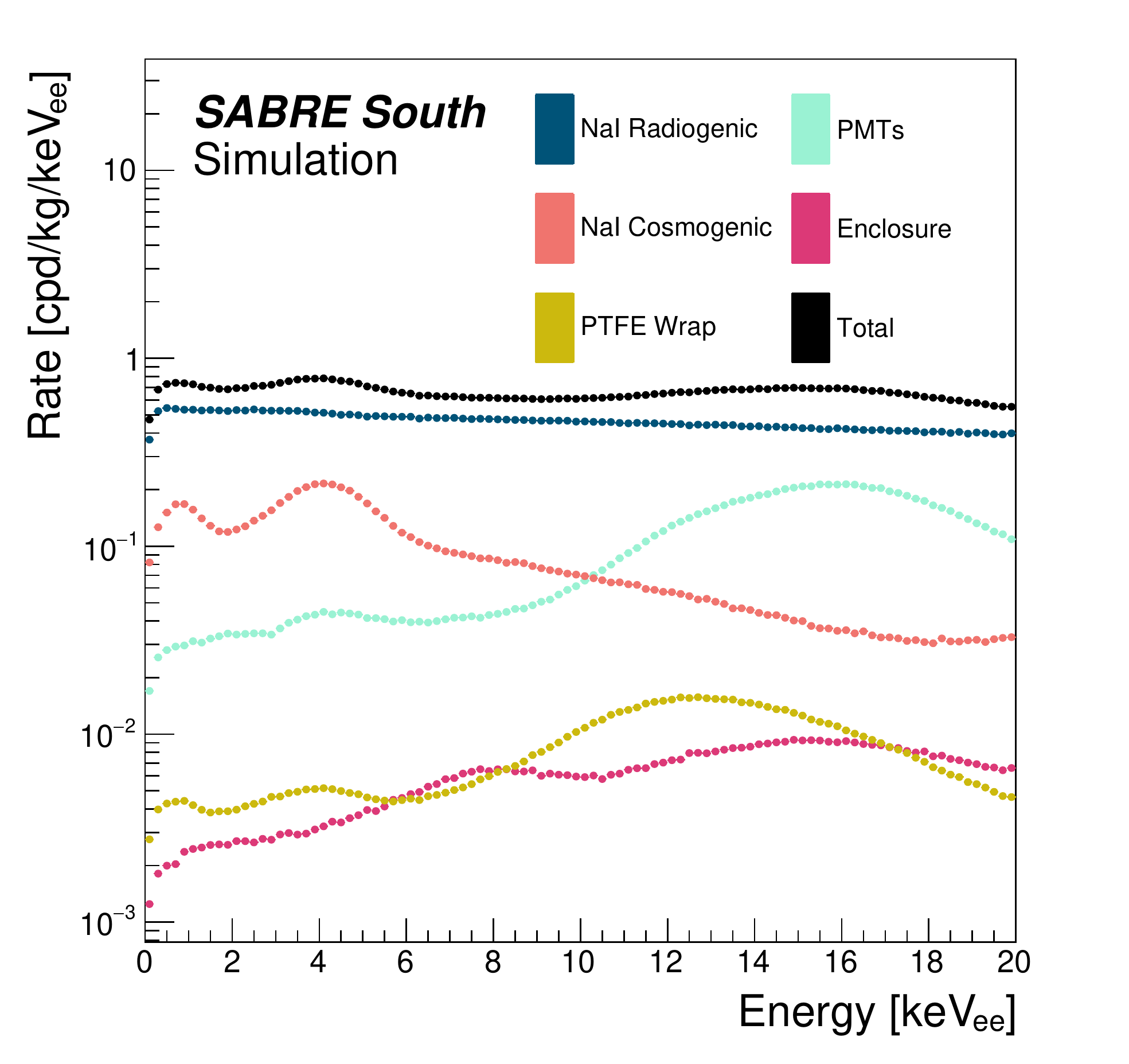}
  \caption{Crystal energy distribution in the range 0--20~\kev of the background from the SABRE South detector components after a 6 month cool-down period. The plot shows contributions down to $10^{-3}~\dru$, which are solely due to materials within the crystal modules. Other components with lower background rate are given in \tab{bulk_dmm_total}. The sum of all the background components is shown in solid black.}
 \label{fig:DMM180_lowEne}
\end{figure}

The background contributions from the SABRE South components to the crystal energy spectrum in the \dmm region are shown in \fig{fig:DMM180_lowEne} and the integrated rates are given in \tab{bulk_dmm_total}. The table also shows the fraction of background suppressed by the veto requirement.

\begin{table}[htb!]
\footnotesize
\centering
\begin{tabular}{lcc}
\hline
 & Rate & Veto Efficiency\\
 & [ \dru]   & [\%]\\
\hline
Crystal radiogenic & $5.2 \cdot 10^{-1}$ & 13 \\
Crystal cosmogenic & $1.6 \cdot 10^{-1}$ & 40 \\
Crystal PMTs & $3.8 \cdot 10^{-2}$ & 60 \\
PTFE wrap & $4.5 \cdot 10^{-3}$ & 13 \\
Enclosures & $3.2 \cdot 10^{-3}$ & 85 \\
Conduits & $1.9 \cdot 10^{-5}$ & 96 \\
Liquid scintillator & $4.9 \cdot 10^{-8}$ & $>99$\\
Steel vessel & $1.4 \cdot 10^{-5}$ & $>99$ \\
Veto PMTs & $1.9 \cdot 10^{-5}$ & $>99$ \\
Shielding & $3.9 \cdot 10^{-6}$ & $>99$ \\
External & O($10^{-4}$) & $>99$ \\
\hline
Total & $7.2 \cdot 10^{-1}$ & 27 \\
\hline
\end{tabular}
  \caption{Background rate in the \dmm region for the SABRE South components after a 6 month cool-down period, and the corresponding veto efficiency.}
  \label{bulk_dmm_total}
\end{table}

The most significant contributions to the background come from contamination in the crystals themselves, accounting for about 95\% of the total $7.2 \cdot 10^{-1}~\dru$. The second-largest contribution comes from the components touching the crystal surfaces, with the crystal PMTs and PTFE wrapping producing $3.8 \cdot 10^{-2}$ and $4.5 \cdot 10^{-3}~\dru$, respectively. Enclosure components produce a total background of $3.2 \cdot 10^{-3}~\dru$, and the components outside the detector modules contribute less than $10^{-4}~\dru$ to the total rate. These contributions are out of range in \fig{fig:DMM180_lowEne}, but their average rates in the \dmm region are reported in~\tab{bulk_dmm_total}.
In general, the further the component from the crystal, the lower the contribution to the background. Conversely, the veto rejection efficiency increases with distance as radiation passes through more of the liquid scintillator medium. Background from external radiation is effectively limited by the active veto and the passive shielding. External gamma radiation is expected to produce order of $10^{-4}~\dru$ background rate, while thermal neutrons are estimated to be around $10^{-7}~\dru$ and the fast neutron background $10^{-8}$ \dru. It is not possible to give a more precise estimate at this stage, due to the approximations adopted in the calculation described in \sect{sect:ext_rad}.

\begin{figure}[t!] \centering
  \includegraphics[width=0.46\textwidth]{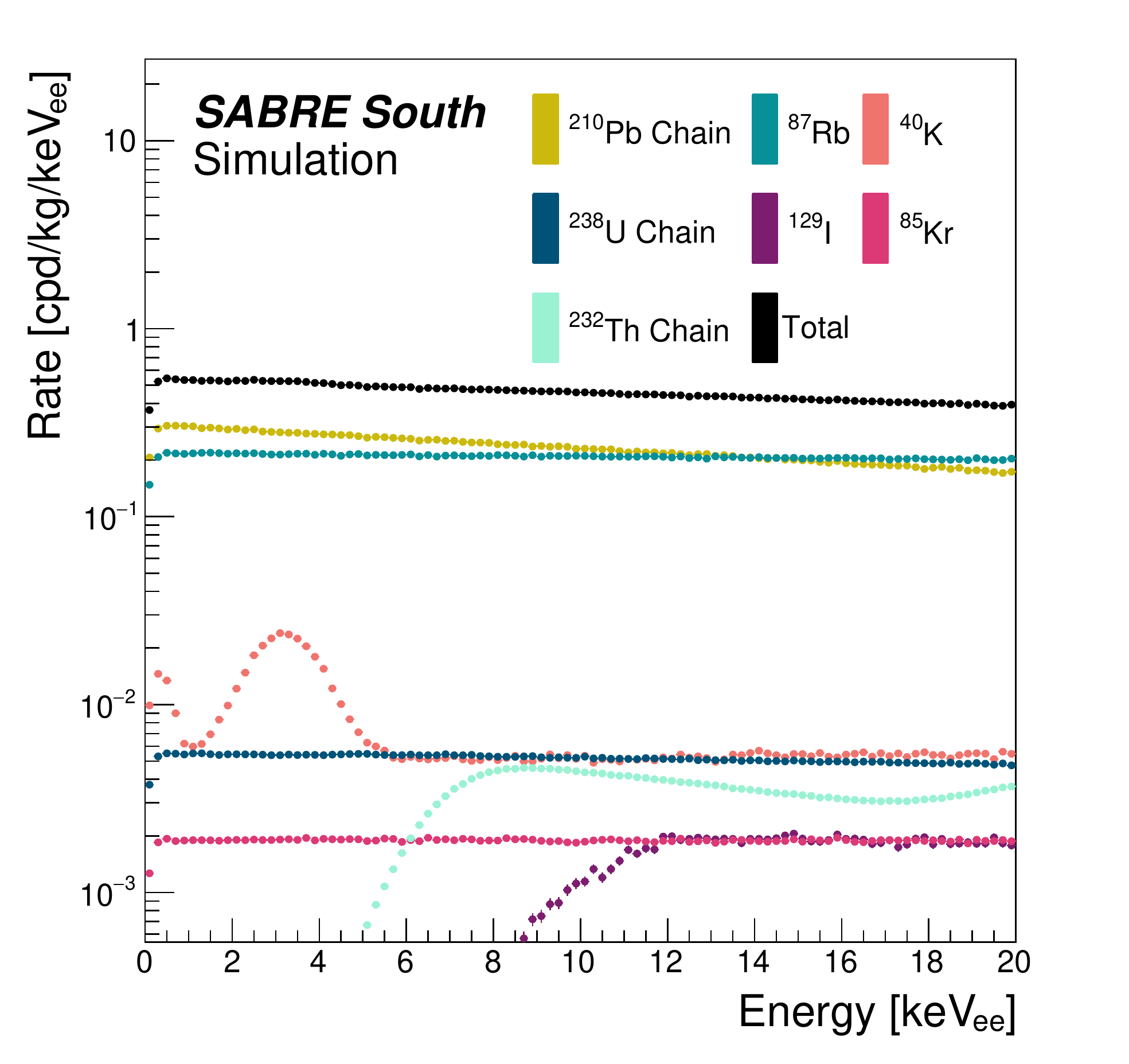}
  \includegraphics[width=0.46\textwidth]{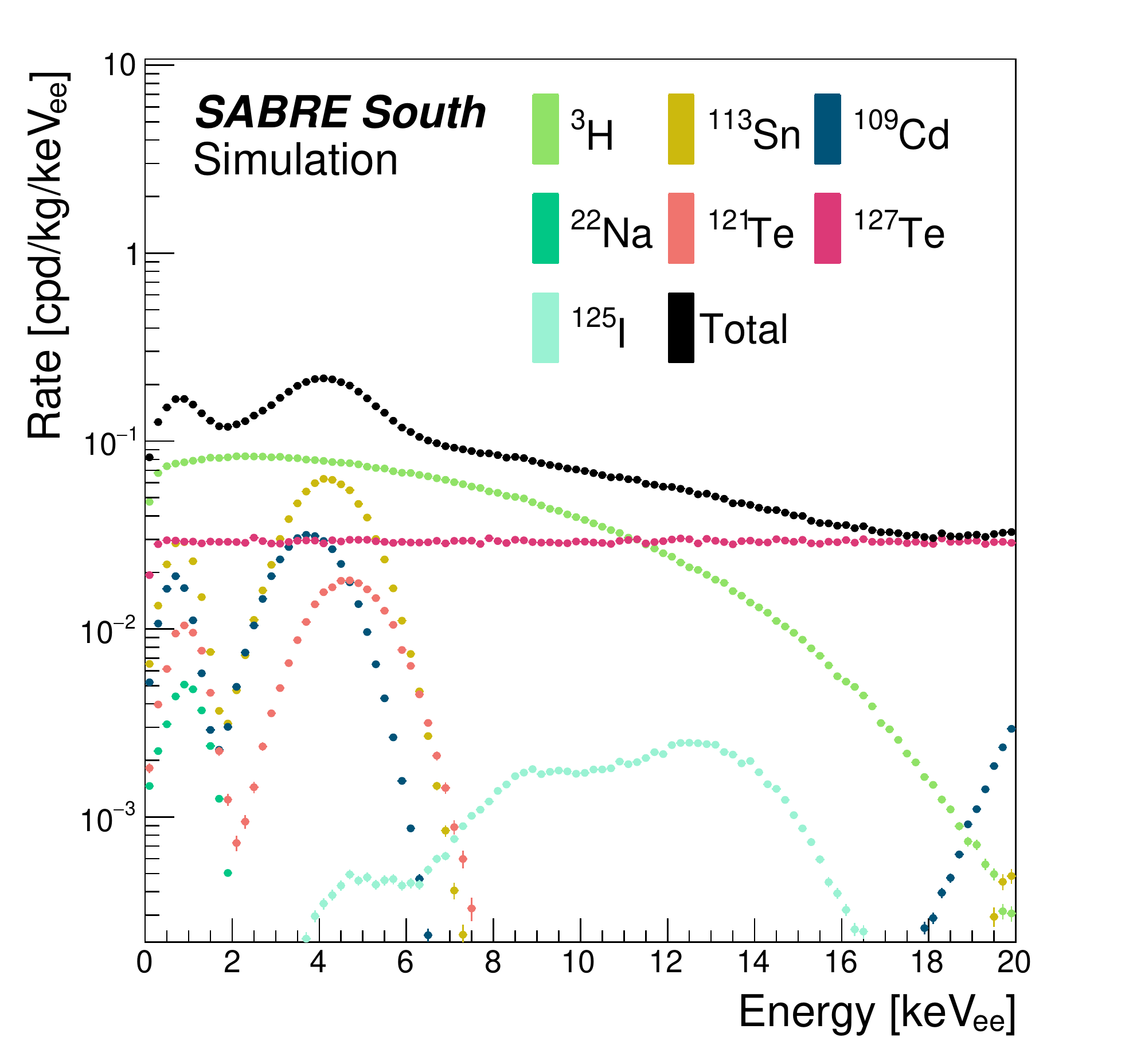}
  \caption{Crystal energy distribution in the range 0--20~\kev of the background due to radiogenic (top) and cosmogenic (bottom) contaminations in the \naitl crystals after a 6 month cool-down period.
The sums of the individual components are also shown (black).}
  \label{fig:crystal_lowEne}
\end{figure}

The background from individual radioisotopes in the crystals is shown in \fig{fig:crystal_lowEne}, with the corresponding rate given in \tab{bulk_NaI}.
\begin{table}[htb!]
\footnotesize
\centering
\begin{tabular}{ccc}
\hline
\multicolumn{3}{c}{Radiogenic} \\
\hline
Isotope & Rate, veto ON & Rate, veto OFF\\
        & [ \dru]   & [ \dru]\\
\hline
  $^{210}$Pb &  $2.8 \cdot 10^{-1}$ & $2.8 \cdot 10^{-1} $ \\
  $^{87}$Rb  &  $<2.2 \cdot 10^{-1}$ & $<2.2 \cdot 10^{-1} $ \\
  $^{40}$K   &  $1.3 \cdot 10^{-2}$ & $1.0 \cdot 10^{-1}$ \\
  $^{238}$U  &  $<5.4 \cdot 10^{-3}$ & $<5.7 \cdot 10^{-3}$ \\
  $^{85}$Kr &   $<1.9 \cdot 10^{-3}$ & $<1.9 \cdot 10^{-3} $ \\
  $^{232}$Th &  $<3.4 \cdot 10^{-4}$ & $<3.9 \cdot 10^{-4} $ \\
  $^{129}$I &   $9.2 \cdot 10^{-5}$ & $9.2 \cdot 10^{-5} $ \\
\hline
  Total & $<5.2 \cdot 10^{-1}$ & $<6.0 \cdot 10^{-1}$  \\
\hline
\multicolumn{3}{c}{Cosmogenic}\\
\hline
Isotope & Rate, veto ON & Rate, veto OFF\\
        & [ \dru]   & [ \dru]\\
\hline
  $^{3}$H        &    $7.8 \cdot 10^{-2}$   &  $7.8 \cdot 10^{-2}$  \\ 
  $^{113}$Sn     &    $3.0 \cdot 10^{-2}$   &  $3.0 \cdot 10^{-2}$  \\ 
  $^{127}$Te     &    $2.9 \cdot 10^{-2}$   &  $2.9 \cdot 10^{-2}$  \\ 
  $^{109}$Cd     &    $1.4 \cdot 10^{-2}$   &  $1.4 \cdot 10^{-2}$  \\ 
  $^{121}$Te     &    $9.1 \cdot 10^{-3}$   &  $1.0 \cdot 10^{-1}$  \\ 
  $^{22}$Na      &    $5.2 \cdot 10^{-4}$   &  $1.4 \cdot 10^{-2}$  \\ 
  $^{125}$I      &    $2.3 \cdot 10^{-4}$   &  $2.3 \cdot 10^{-4}$  \\ 
  $^{113m}$In    &    $7.5 \cdot 10^{-5}$   &  $5.2 \cdot 10^{-4}$  \\ 
  $^{127m}$Te    &    $4.9 \cdot 10^{-5}$   &  $4.9 \cdot 10^{-5}$  \\ 
  $^{126}$I      &    $4.1 \cdot 10^{-5}$   &  $6.2 \cdot 10^{-5}$  \\ 
  $^{121m}$Te    &    $1.8 \cdot 10^{-5}$   &  $6.0 \cdot 10^{-5}$  \\ 
  $^{123m}$Te    &    $7.3 \cdot 10^{-6}$   &  $1.3 \cdot 10^{-5}$  \\ 
  $^{109m}$Ag    &    $2.8 \cdot 10^{-6}$   &  $2.8 \cdot 10^{-6}$  \\ 
  $^{125m}$Te    &    $1.6 \cdot 10^{-6}$   &  $1.7 \cdot 10^{-6}$  \\ 
\hline                     
  Total    &   $1.6 \cdot 10^{-1}$    &  $2.7 \cdot 10^{-1}$ \\ 
\hline
\end{tabular}
\caption{Background rate in the \dmm region due to the contaminants in NaI(Tl) crystals. Both radiogenic and cosmogenic contributions are reported with and without veto requirement. The contributions are listed in decreasing order with veto on. The $<$ sign indicates upper limits for isotopes that were not detected and are limited by the screening measurement sensitivity. Cosmogenic backgrounds are computed after a 180 days cool-down period underground.}
\label{bulk_NaI}
\end{table}
Among the radiogenic contamination, the highest contributions come from $^{210}$Pb ($2.8 \cdot 10^{-1}~\dru$) and $^{87}$Rb ($<2.2 \cdot 10^{-1}~\dru$). The $^{87}$Rb contribution, however, is an upper limit dictated by experimental precision. No $^{87}$Rb was found with the ICP-MS measurement, and the order of magnitude of this contamination is currently unknown. The third most relevant contribution is $^{40}$K but this is efficiently suppressed by the veto requirement down to $1.3 \cdot 10^{-2}~\dru$. Isotopes in the $^{238}$U chain are responsible for $5.4 \cdot 10^{-3}~\dru$, while the $^{232}$Th chain gives $3.4 \cdot 10^{-4}~\dru$.

\begin{figure}[htb!] \centering
  \includegraphics[width=0.46\textwidth]{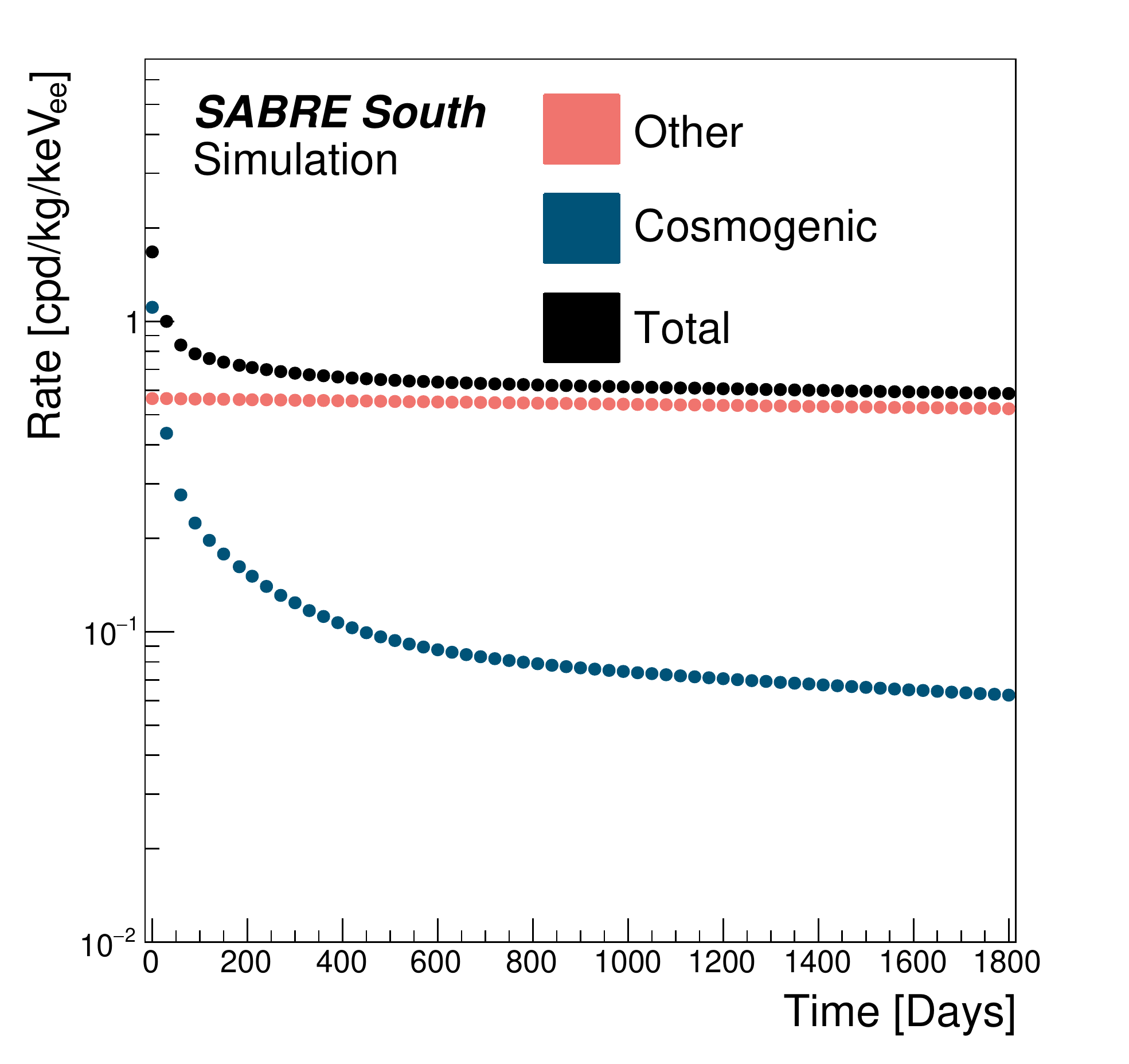}
  \caption{Time-dependent background rate in the \dmm region. The total rate is shown in black, the radiogenic component in orange and the long-lived background in blue. The origin corresponds to the time of first placement of the crystals underground. }
  \label{fig:bkg_vs_time}
\end{figure}

The background due to cosmogenic activation in crystals is dominated by $^{3}$H, $^{113}$Sn, $^{127}$Te and $^{109}$Cd based on the crystal exposure history and cooling off detailed in \sect{sec:cont_crystal}. 
Individual radioisotopes will indeed contribute differently to the background over the lifetime of the experiment due to their different half-lives. $^{3}$H has the longest half-life (4497 days) and will produce a nearly constant background for years. $^{113}$Sn and $^{109}$Cd have instead hundreds of days half-life and will decrease more rapidly. We expect these isotopes to contribute to the background $10^{-6}$ and $10^{-3}~\dru$ respectively after five years. Some radioisotopes, such as $^{127}$Te, decay even faster but they can be regenerated by the decay of other isotopes ($^{127m}$Te for $^{127}$Te). We have taken into account decay and production rates of radioisotopes once the detector is underground and calculated the expected total time-dependent background rate of the experiment. \fig{fig:bkg_vs_time} shows the background rate in the \dmm region as a function of time from the placement of the crystals underground, i.e. once cosmic-ray exposure has ceased. The cosmogenic background is expected to be at the same level of other long lived isotopes after 20 days of cool-down and five times smaller after one year.

\begin{figure}[htb!] \centering
  \subfigure[Total background.]
  {\includegraphics[width=0.46\textwidth]{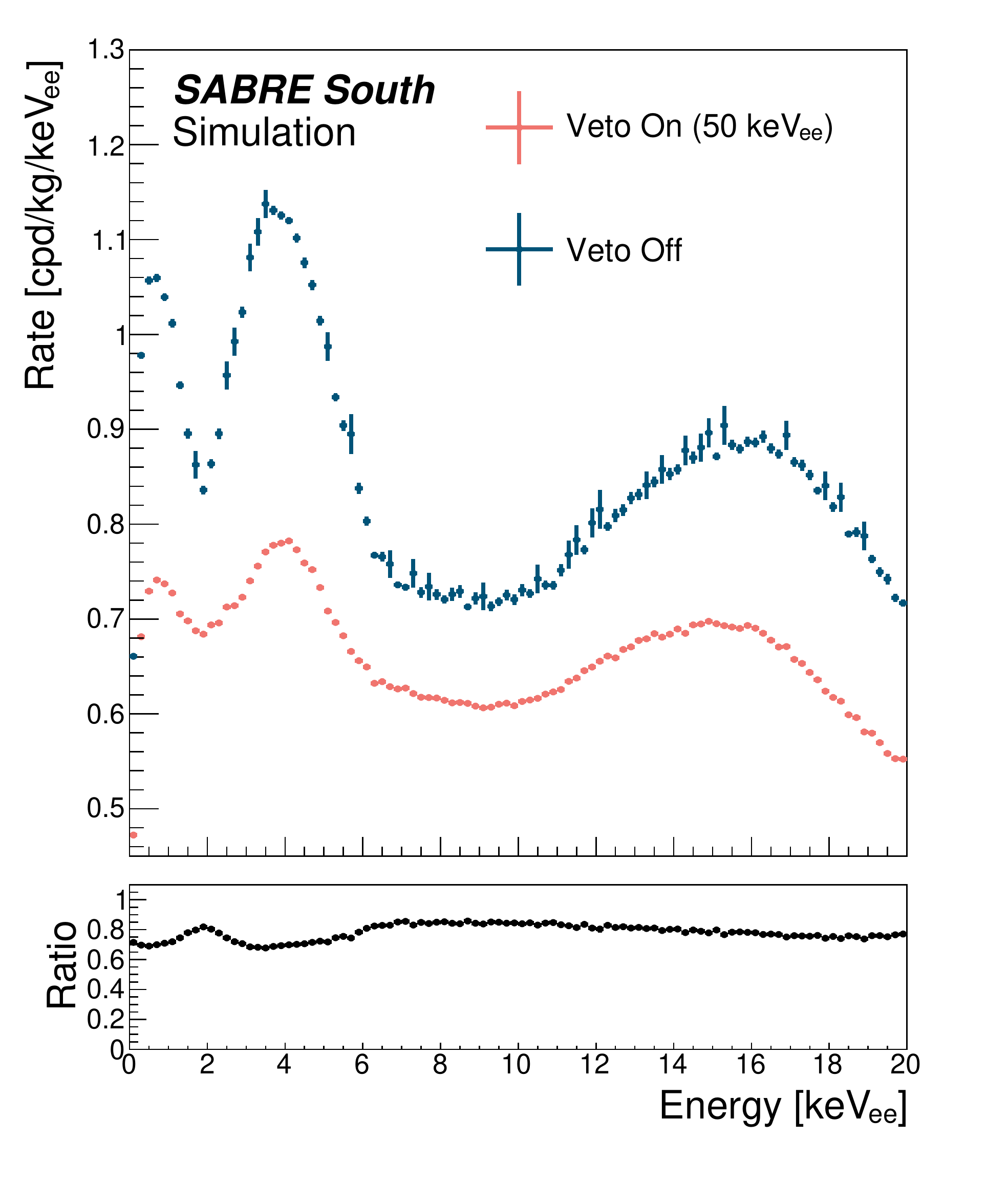}}
   \hspace{-5mm}
  \subfigure[Crystal $^{40}$K background.]
  {\includegraphics[width=0.46\textwidth]{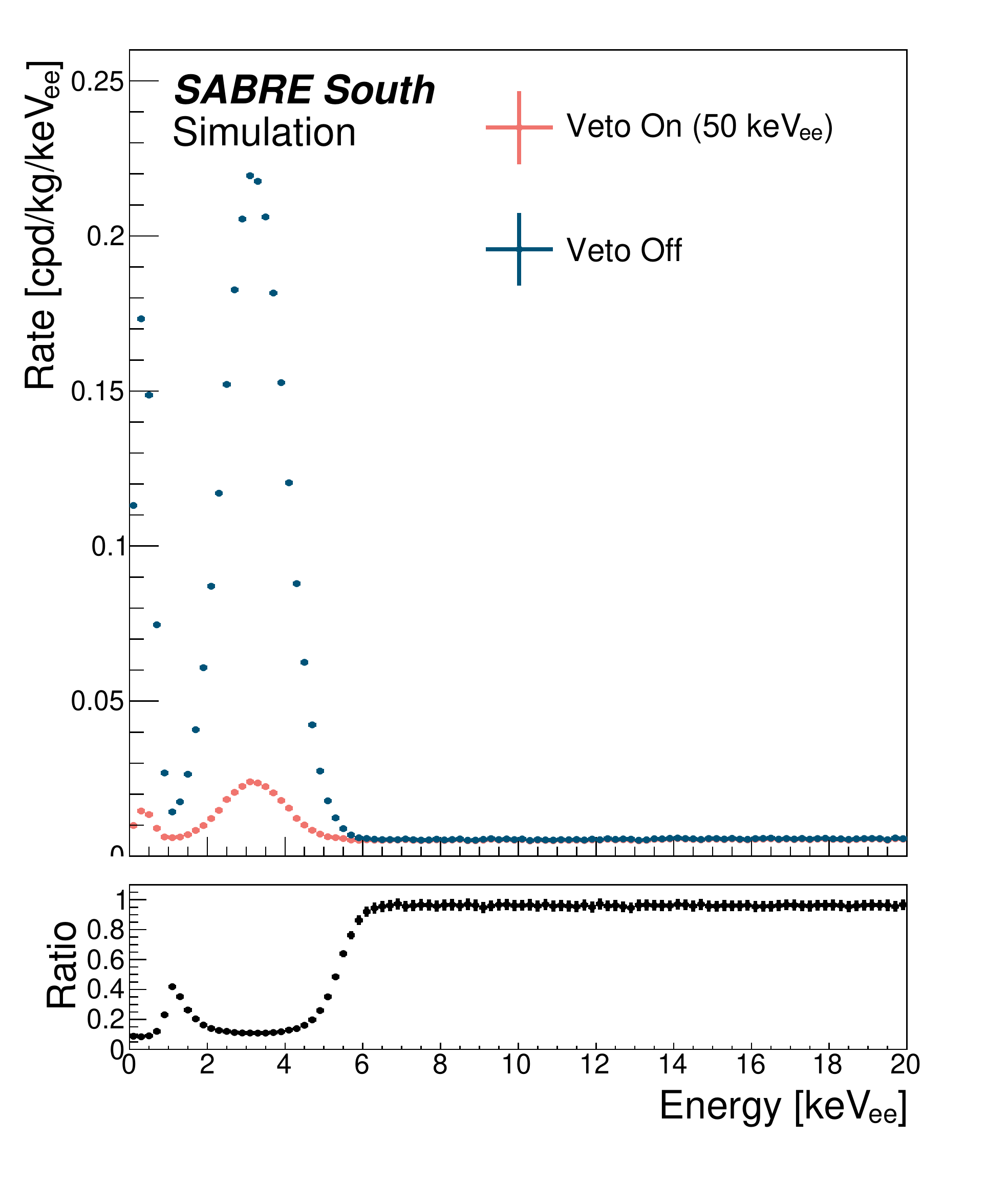}}
   \hspace{-5mm}
  \caption{Crystal energy distribution in the range 0--20~\kev of the total background (a) and of the component due to $^{40}$K decays in the crystal (b). The spectra are shown with (orange) and without (blue) the veto requirement. The veto reduction factor is given as a function of energy in the lower panel.}
  \label{fig:VETOOnOff}
\end{figure}

Overall, the veto requirement is expected to suppress 27\% of the total background. The efficacy of the veto relies on the presence of high-energy decay products that can escape the crystals. As a number of key backgrounds (such as $^{210}$Pb and $^{3}$H) lack this feature, they cannot be vetoed. When penetrating radiation is emitted, as in the case of $^{40}$K, the veto efficiency increases significantly. For $^{40}$K, electron capture occurs with an 11\% branching ratio, producing a 3.2~keV X-ray or Auger de-excitation in coincidence with a 1.46~MeV $\gamma$. The high-energy $\gamma$ can escape the crystal and be detected in the liquid scintillator or in another crystal with an expected efficiency of about 87\%. %\fig{fig:KVETO} shows the background spectrum of $^{40}$K decays in the crystal with and without the veto requirement.
\fig{fig:VETOOnOff} shows the crystal energy distributions with and without the veto requirement of the total background and of the $^{40}$K component.
% \begin{figure}[htb!] \centering
%   \includegraphics[width=0.46\textwidth]{Veto_ONOFF_comparison_40K_CrystalEnergySmear0-20.pdf}
%   \caption{Crystal energy distribution in the range 0--20~\kev of the background due to $^{40}$K decays in the crystal with (orange) and without (blue) the veto requirement. The bottom panel shows the veto reduction factor as a function of energy.}
%   \label{fig:KVETO}
% \end{figure}
% \begin{figure}[htb!] \centering
%   \includegraphics[width=0.46\textwidth]{Veto_ONOFF_comparison_Total_CrystalEnergySmear0-20.pdf}
%   \caption{Crystal energy distribution in the range 0--20~\kev of the total background with (orange) and without (blue) the veto requirement. The bottom panel shows the veto reduction factor as a function of energy.}
%   \label{fig:TotalVETO}
% \end{figure}
The active veto system also reduces very efficiently the $^{121}$Te background, which would otherwise account for $0.1~\dru$ even after six months of cool-down. 
It is worth noting that this system also plays a fundamental role in suppressing external radiation as the steel-polyethylene passive shielding alone is not capable of stopping enough radiation. In this scenario we would indeed expect a contribution to the background from external radiation of the order of 1 \dru. Thus, the active veto system is an essential feature of the SABRE South experiment, which both lowers the background from the crystal and suppresses the external radiation background to a negligible level.

\begin{figure}[htb!] \centering
  \includegraphics[width=0.46\textwidth]{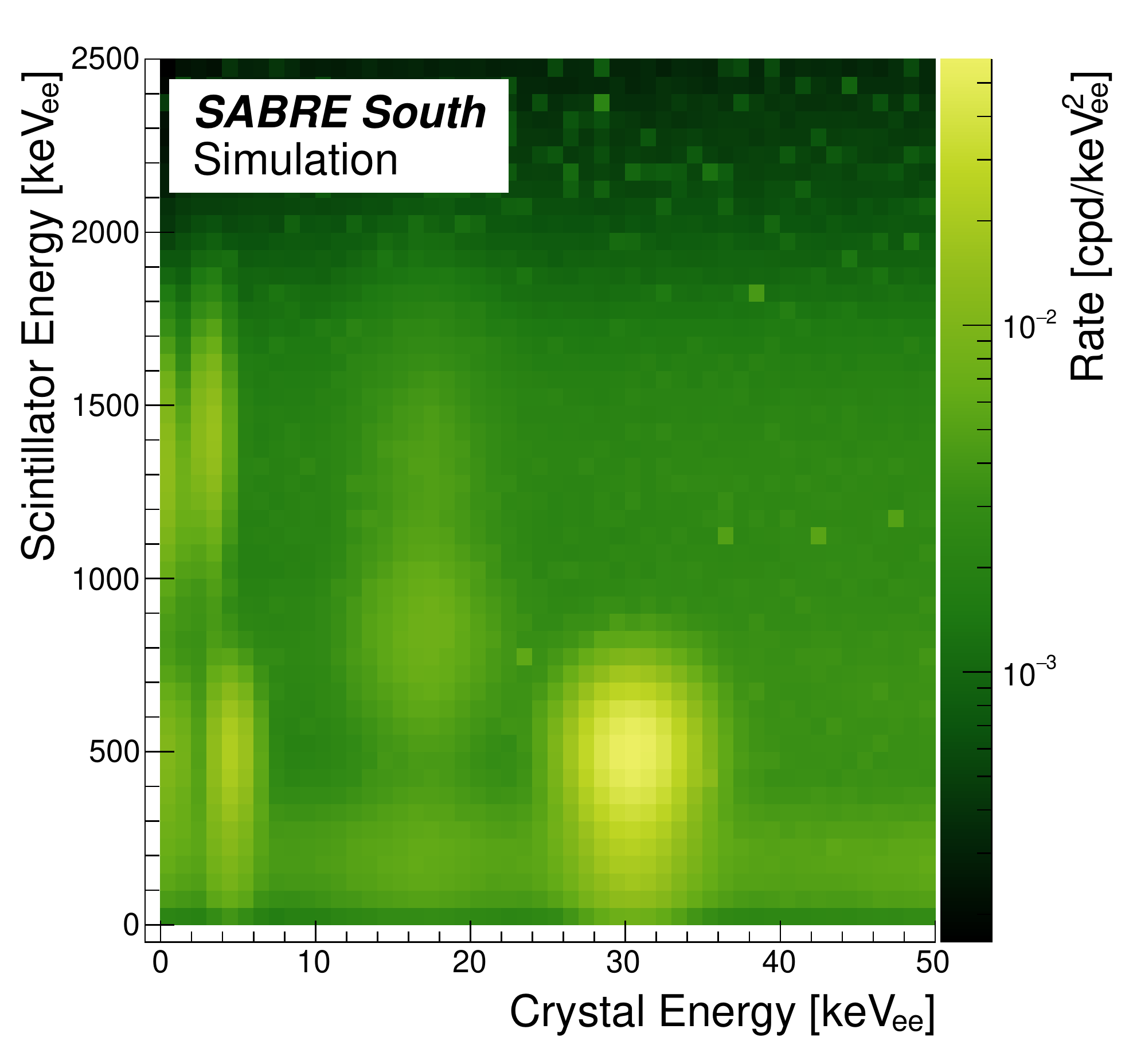}
  \caption{Background rate distribution in events with crystal and liquid scintillator coincidences as a function of detected crystal energy and scintillator energy.}
  \label{fig:crys_vs_scint}
\end{figure}

\begin{figure}[htb!] \centering
  \includegraphics[width=0.46\textwidth]{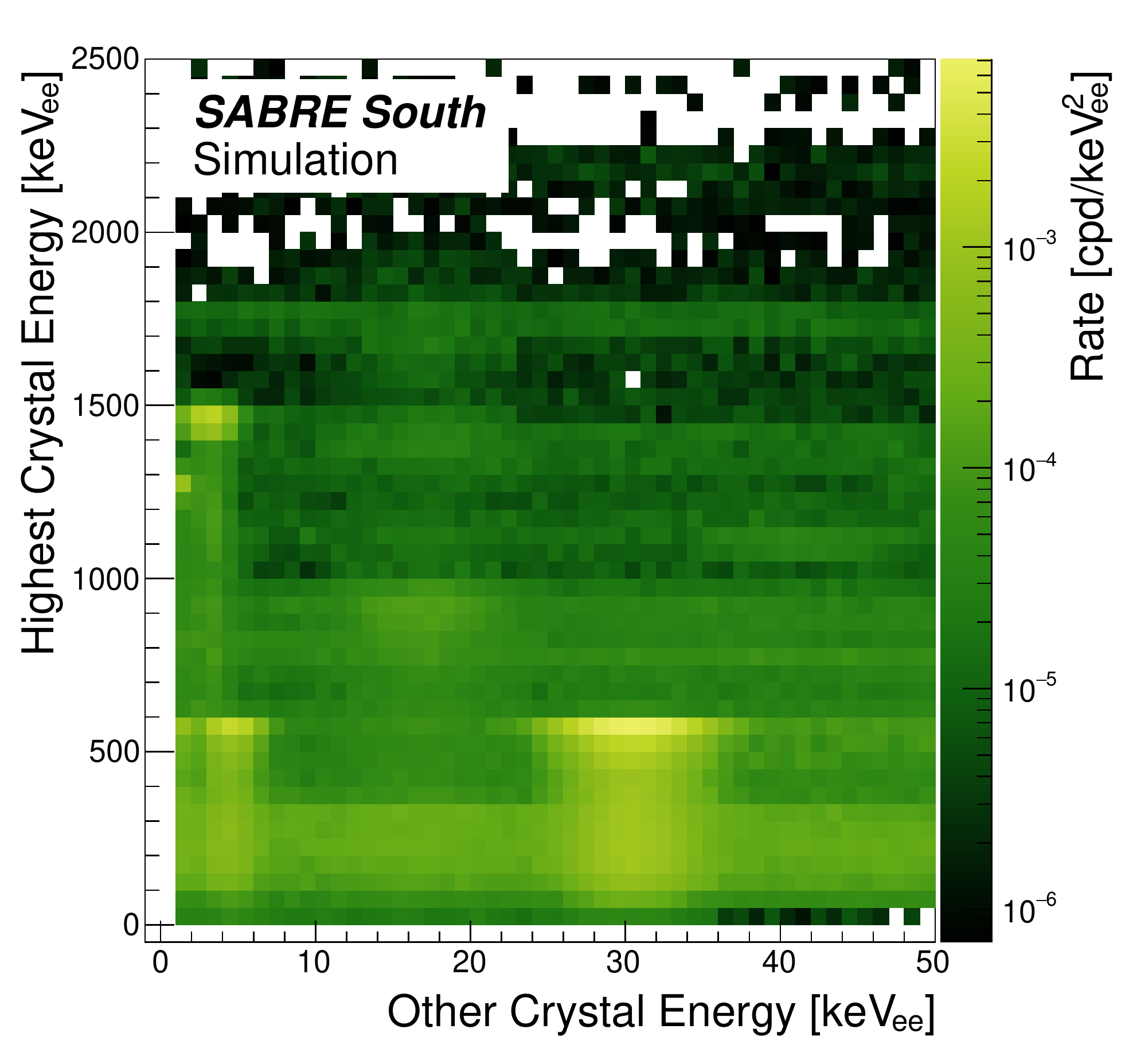}
  \caption{Background rate distribution in events with crystal coincidences as a function of the two highest energy depositions in separate crystals.}
  \label{fig:crys_vs_crys}
\end{figure}

The presence of energetic gamma-ray emissions in $^{40}$K and $^{121}$Te also offers a method of measuring their level of contamination in the crystals. Such an analysis will be performed using coincidences between crystal detectors and/or the liquid scintillator. \fig{fig:crys_vs_scint} shows the energy distributions of background events with energy deposited in one crystal and in the liquid scintillator. The excess centered around 3~\kev of crystal energy and 1.46~\mev of liquid scintillator energy is due to $^{40}$K contamination in the crystal, while those at 5 and 30~\kev crystal energy and 570~\kev scintillator energy are due to $^{121}$Te. These excesses can also be observed in events with coincidences between two crystals and with no energy in the veto above 50~\kev at the same energy values as shown in \fig{fig:crys_vs_crys}. The $^{40}$K and $^{121}$Te spread onto a wider area in~\fig{fig:crys_vs_scint} due to the poorer resolution of the liquid scintillator compared to the crystal. Using \figs{fig:crys_vs_scint}{fig:crys_vs_crys}, we have defined preliminary regions for the measurement of $^{40}$K and $^{121}$Te activities. The $^{40}$K measurement region A and the $^{121}$Te measurement region A are designed to collect events where the higher-energy gamma-ray from these decays is detected in the liquid scintillator. Events where the gamma-ray is detected in a \naitl detector fall into in the $^{40}$K measurement region B and the $^{121}$Te measurement region B, instead. The selection requirements defining these regions are given in \tab{K40_Te121_sel}. The expected detection rate and the sample purity in each region is reported in~\tab{K40_Te121_meas}. If the $^{40}$K and $^{121}$Te activities are similar to those assumed in this background model (see \tab{tab:intrinsic-rad} and \tab{tblbulk_NaI}) a direct measurement of them in this way should be feasible with just few months of data.

\begin{table*}[hbt!]
\footnotesize
\centering
\begin{tabular}{lllll}
\hline
 & \multicolumn{2}{c}{$^{40}$K measurement}  & \multicolumn{2}{c}{$^{121}$Te measurement}\\
Selection & \multicolumn{1}{c}{Region A} & \multicolumn{1}{c}{Region B} & \multicolumn{1}{c}{Region A} & \multicolumn{1}{c}{Region B} \\
\hline
Liquid Scintillator Energy [\kev] & $E\in[1000,1800]$  & $E<50$ & $E\in[150,800]$  & $E<50$ \\
Highest \naitl Energy [\kev] & $E\in[2,4]$ & $E\in[1400,1500]$ & $E\in[26,35]$ & $E\in[450,600]$ \\
$2^{nd}$ Highest \naitl Energy [\kev] &  $E<1$ & $E\in[1,5]$ & $E<1$ & $E\in[25,36]$ \\
\hline
\end{tabular}
\caption{Definition of the $^{40}$K and $^{121}$Te measurement regions.}
\label{K40_Te121_sel}
\end{table*}
\begin{table}[htb!]
\footnotesize
\centering
\begin{tabular}{ccc}
\hline
Region & Total rate [cpd] & Sample purity\\
\hline
$^{40}$K measurement region A & 13 & $90\%$\\
$^{40}$K measurement region B & 0.32 & $>99\%$\\
$^{121}$Te measurement region A & 130 & $90\%$\\
$^{121}$Te measurement region B & 3.2 & $>99\%$\\
\hline
\end{tabular}
\caption{Expected event rate and sample purity in the $^{121}$Te and $^{40}$K measurement regions.}
\label{K40_Te121_meas}
\end{table}

\subsection{Projected sensitivity of the SABRE South experiment}
\begin{figure}[htb!]
    \centering
    \includegraphics[width=0.5\textwidth]{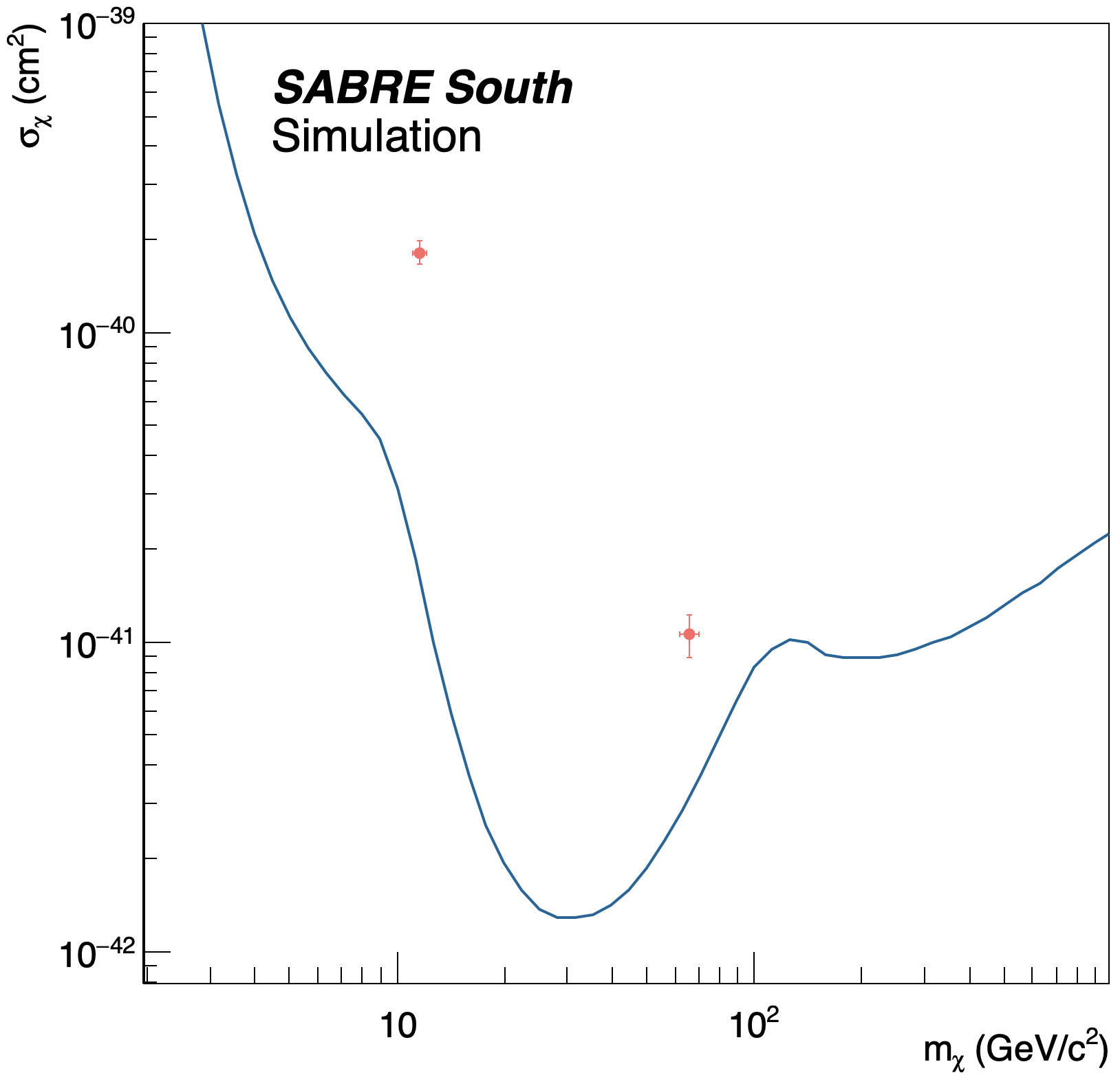}
    \caption{90\% exclusion curve for the SABRE South experiment after three years of data taking (in blue) assuming a background model in the 1--6~\kev region given by \fig{fig:DMM180_lowEne} and an exposure mass of 50 kg. The best fits to the \dama data for this model in both the low- and high-mass region are shown in pink.}
    \label{fig:sensitivity}
\end{figure}

The sensitivity of SABRE South to a typical WIMP has also been computed. These calculations are performed assuming the spin-independent effective field theory operator $\mathcal{O}1$ from Ref. \cite{Fitzpatrick_2013}, Standard Halo Model velocity distribution, an efficiency equivalent to that of COSINE \cite{coseff}, and the \dama quenching factor values for both Na and I. \fig{fig:sensitivity} shows the 90\% confidence level (CL) limit obtained using the method detailed in Ref. \cite{Zurowski_2021} assuming 50 kg of target mass, three-year exposure and the constant background energy spectrum given by \fig{fig:DMM180_lowEne} in the 1--6~\kev region. The SABRE South data acquisition system is capable of recording events at a rate of several orders of magnitude higher than the expected crystal signal rate, resulting in zero dead-time. We expect to lose less than $1\%$ of processes producing signals in crystals due to accidental coincidences in the veto.
For this model we include (for comparison) the best fits to \dama~\cite{DAMAPhase2} in both the low-mass (preferential Na coupling) and high-mass (preferential I coupling) regimes, which are also reported in Table \ref{tab:dama_fits}. 
SABRE South should be capable of excluding a signal ten times smaller than the \dama fit for the low-mass and four times smaller for the high-mass regimes. 
The strongest limit is set at $m_{\chi}=30$ GeV/c$^2$ with cross sections larger than $\sigma_{\chi}=1.04\times 10^{-42}$ cm$^2$ excluded.

\begin{table}[htbp!]
\footnotesize
\centering
\begin{tabular}{ccc}
\hline
 & \multicolumn{2}{c}{$\sigma_{\chi}$ [cm$^2$]}\\ 
\cline{2-3}
$m_{\chi}$ [GeV/c$^2$] & \dama & SABRE South \\
 & fit & limit \\
\hline
$11.6\pm 0.5$        & $1.8\pm 0.2 \cdot 10^{-40}$ & $1.6 \cdot 10^{-41}$   \\
$65.9\pm 4.8$        & $1.1\pm 0.2 \cdot 10^{-41}$ & $3.1 \cdot 10^{-42}$    \\ \hline
\end{tabular}
\label{tab:dama_fits}
\caption{Best fits to the \dama data \cite{DAMAPhase2} for the spin-independent $\mathcal{O}1$~\cite{Fitzpatrick_2013} in the low- and high-mass regions and SABRE South's exclusion limits after three years of exposure and 50~kg of target mass.
}
\end{table}

\begin{figure}[htb!]
    \centering
    \includegraphics[width=0.5\textwidth]{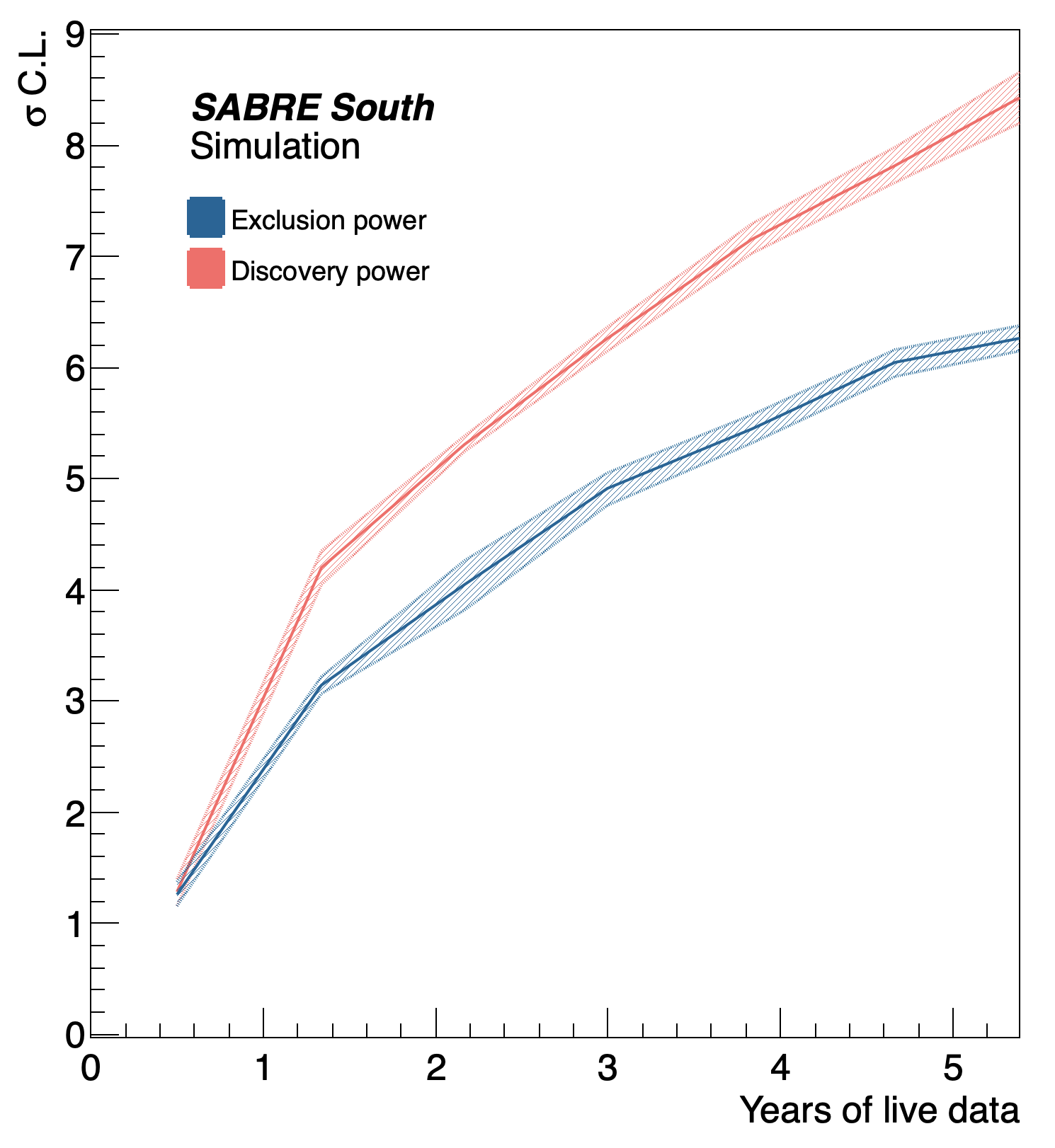}
    \caption{The exclusion and discovery power of SABRE South for a \dama-like signal. The shaded regions indicate 1$\sigma$ statistical uncertainty bands.}
    \label{fig:dama_sensitivity}
\end{figure}

We also estimate the power of SABRE South to exclude or confirm the \dama annual modulation signal~\cite{DAMAPhase2}, assuming a background constant in time. This assesses the ability of SABRE South to observe a signal with a modulation of 0.0119 \dru, following the methodology of \cite{Zurowski_2021}, and is shown in \fig{fig:dama_sensitivity}. Based on these results, with 1.3 (3.0) annual cycles of data, SABRE South will be able to refute the interpretation of the \dama modulation as a dark matter signal with 3$\sigma$ (5$\sigma$) CL. In the event of observation of the annual modulation, this signal would reach a significance of 5$\sigma$ CL with two full years of data. Such rapid results are made possible thanks to the small background rate compared to other experiments, which comes from using ultra-pure crystals and an active veto. If we assume a background of 2~\dru, a 3$\sigma$ exclusion or a 5$\sigma$ observation would require 4-5 years of data-taking. We have also considered the case where we reduce the energy range of analysis to 2--6~\kev. In this scenario, 3$\sigma$ exclusion and 5$\sigma$ observation are expected in about 1.5 and 2.8 years of data-taking, respectively.
As well as this, the sensitivity of SABRE is strongly dependent on the quenching factor value, particularly in the model dependent scenario. Recent results show that the Na QF is typically lower than that reported by \dama \cite{bignell2021,PhysRevC.92.015807,JOO201950}, which would mean that the interpretation of the \dama signal is shifted to larger masses \cite{Ko_2019}. This shifts the sensitivity as a function of mass by a similar factor, and so provided SABRE South and \dama have the same quenching factors, a change in their values will impact only the interpretation of the \dama signal, rather than the exclusion power of SABRE South. 

\section{Conclusions}
\label{sec:conclusion}
We have evaluated the expected background of the SABRE South experiment due to radioactive emissions from the detector components and the external environment. This prediction is based on a \geant simulation of the experiment combined with measurements (or assumptions) of the radiation levels within the detector materials. The simulation carefully reproduces the design of the apparatus, with particular attention to the parts close to the crystal detector.   

We find that the contamination of the crystals gives the most significant contribution to the radioactive background in the \dmm region ($95\%$ of the overall background), confirming the importance of lowering the crystal contamination as much as possible. The radio-purity of the crystals, combined with the active veto technique, allows SABRE to achieve a background of 0.72~\dru in the 1--6~\kev energy region, where the modulation signal was observed by \dama. This background rate includes a 0.22 \dru contribution from the upper limit estimate of $^{87}$Rb in the crystals. No $^{87}$Rb contamination has been reported by other \naitl experiments so far, thus the total background rate for the SABRE South experiment might be lower than what is reported here. 

The dominant contribution within the crystals is expected to be from bulk contamination of $^{210}$Pb in the crystals (0.28 \dru), followed by production of $^{3}$H in the crystals during exposure to cosmic rays ($7.8 \cdot 10^{-2}$ \dru). Based on this simulated background, which does not include PMT noise, SABRE South is expected to reject the \dama modulation at $4\sigma$ (in the case of null results) or confirm it at $5\sigma$ (in the event of observation of a compatible modulation) within 2.5 years.

% -----------------------------------------------------------------
\section*{Acknowledgements}
The SABRE South program is supported by 
the Australian Government through the Australian Research Council (Grants:
CE200100008, 
LE190100\-196, 
LE170100162, 
LE160100080, 
DP190103123, 
DP170101675, 
LP150100705).
This research was partially supported by Australian Government Research Training Program Scholarships and Melbourne Research Scholarships.
This research was supported by The University of Melbourne’s Research Computing Services and the Petascale Campus Initiative.
We thank the SABRE North collaboration for their contribution to the SABRE South experiment design and to the simulation framework.
We also thank the Australian Nuclear Science and Technology Organisation for the assistance with the material screening and the measurement of background radiation at SUPL.

% -----------------------------------------------------------------

\bibliographystyle{spphys}
\interlinepenalty=10000
\bibliography{main}
\end{document}